\documentclass[
  aps,
  amsfonts,
  amsmath,
  amssymb,
  prb,
  twocolumn,
  superscriptaddress,
  longbibliography,
  10pt
]{revtex4-2}

\usepackage[utf8]{inputenc}
\usepackage{bbm}
\usepackage{graphicx}
\usepackage[dvipsnames]{xcolor}
\usepackage[colorlinks=true]{hyperref}
\usepackage{orcidlink}
\usepackage{bm}
\usepackage{slashed}
\usepackage{tabularray}
\usepackage{siunitx}
\usepackage[mathscr]{euscript}
\usepackage[section]{placeins}
\usepackage{empheq}
\usepackage{enumitem}

\hypersetup{
    colorlinks=true,
    linkcolor=magenta,
    citecolor=blue,
    filecolor=magenta,
    urlcolor=blue
}

\newcommand{\mbf}[1]{\mathbf{#1}}
\newcommand{\dd}{\mathrm{d}}

\newcommand{\vi}{{\boldsymbol{i}}}
\newcommand{\vj}{{\boldsymbol{j}}}
\newcommand{\vecr}{{\boldsymbol{r}}}
\newcommand{\vk}{{\boldsymbol{k}}}
\newcommand{\vK}{{\boldsymbol{K}}}
\newcommand{\vb}{{\boldsymbol{b}}}

\newcommand{\vq}{{\boldsymbol{q}}}

\newcommand{\SU}{\mathrm{SU}}
\newcommand{\U}{\mathrm{U}}
\newcommand{\Z}{\mathbb{Z}}

\newcommand{\eqnref}[1]{Eq.~\eqref{#1}}
\newcommand{\figref}[1]{Fig.~\ref{#1}}
\newcommand{\tabref}[1]{Table~\ref{#1}}
\newcommand{\secref}[1]{Sec.~\ref{#1}}
\newcommand{\appref}[1]{Appendix~\ref{#1}}

\begin{document}

\author{Andreas Feuerpfeil\,\orcidlink{0009-0001-0436-0332}}
\email[Contact Author: ]{andreas.feuerpfeil@uni-wuerzburg.de}
\affiliation{Institut für Theoretische Physik und Astrophysik and W\"urzburg-Dresden Cluster of Excellence ctd.qmat, Universit\"at W\"urzburg, 97074 W\"urzburg, Germany}
\affiliation{Center for Computational Quantum Physics, Flatiron Institute, 162 5th Avenue, New York, New York 10010, USA}
\author{Atanu Maity\,\orcidlink{0000-0001-7822-124X}}
\affiliation{Institut für Theoretische Physik und Astrophysik and W\"urzburg-Dresden Cluster of Excellence ctd.qmat, Universit\"at W\"urzburg, 97074 W\"urzburg, Germany}
\author{Ronny Thomale\,\orcidlink{0000-0002-3979-8836}}
\affiliation{Institut für Theoretische Physik und Astrophysik and W\"urzburg-Dresden Cluster of Excellence ctd.qmat, Universit\"at W\"urzburg, 97074 W\"urzburg, Germany}
\affiliation{Department of Physics, Indian Institute of Technology Madras, Chennai 600036, India}
\author{Yasir Iqbal\,\orcidlink{0000-0002-3387-0120}}
\affiliation{Department of Physics, Indian Institute of Technology Madras, Chennai 600036, India}
\author{Subir Sachdev\,\orcidlink{0000-0002-2432-7070}}
\email[Contact Author: ]{sachdev@g.harvard.edu}
\affiliation{Department of Physics, Harvard University, Cambridge, Massachusetts 02138, USA}
\affiliation{Center for Computational Quantum Physics, Flatiron Institute, 162 5th Avenue, New York, NY 10010, USA}

\title{Higgs criticality of Dirac spin liquids on depleted triangular lattices}

\date{\today}

\begin{abstract}
We investigate Higgs criticality in candidate U(1) Dirac spin liquids across a family of depleted triangular lattices: the triangular, kagome, and maple-leaf geometries. For each, we identify the symmetry-allowed spinon-pairing channel connecting the U(1) state to a proximate $\mathbb{Z}_2$ spin liquid, deriving the corresponding quantum electrodynamics (QED$_3$)-Higgs theory. While the triangular and kagome lattices share a low-energy description with $N_f=4$ Dirac fermions, the maple-leaf lattice yields an analogous theory with $N_f=12$ and a distinct nodal structure where the Dirac cones can move along high-symmetry lines in momentum space instead of gapping out. Using a large-$N_{f,b}$ expansion, we compute critical exponents and the scaling dimensions of the symmetry-allowed Yukawa couplings. Although Higgs-field fluctuations and a large fermion flavor number strongly suppress the Yukawa coupling---pushing the maple-leaf lattice closer to stability than its counterparts---we find that the coupling remains relevant at leading order in all three cases, substantially reducing its scaling dimension on the maple-leaf lattice and asymptotically breaking Lorentz and conformal invariance. Ultimately, our results provide a unified framework demonstrating how the interplay between fermion flavor count and nodal geometry dictates the fate of the QED$_3$-Higgs transition.
\end{abstract}

\maketitle

\section{Introduction}
\label{sec:intro}

Quantum spin liquids on frustrated lattices remain one of the most fruitful settings in which to examine how continuum gauge theories emerge from microscopic quantum magnets. Among them, a particularly important role is played by the $\U(1)$ Dirac spin liquid, whose long-wavelength description is given by quantum electrodynamics in $2+1$ dimensions (QED$_3$), with massless Dirac spinons coupled to an emergent compact $\U(1)$ gauge field. In recent years, this field-theoretic description has moved well beyond a purely formal construction. Especially on lattices built from triangular motifs, numerical spectra, variational wave functions, and symmetry analyses have made it possible to compare microscopic results directly with the operator content of QED$_3$ and to test this picture in increasingly concrete ways~\cite{Song-2019,Song-2020,Wietek24,Budaraju-2023,Budaraju-2025}. 

Parallel to these theoretical advances, experimental developments have lent these questions additional timeliness. Experiments on organic materials~\cite{Kanoda18,Kanoda24} and NaYbSe$_2$~\cite{Tennant24,chinese_na} show evidence of spin-liquid behavior on triangular lattices, along with proximate superconductivity. More recently, twisted bilayer WSe$_2$ has been shown to host superconductivity near half-band filling together with a displacement-field-tuned continuous transition into a correlated insulating state~\cite{Xia2025,Xia2026}, motivating broader questions about triangular-lattice Mott physics, fractionalization, and pairing~\cite{Kim2024,Peng2025}. We stress, however, that the present work is not a microscopic theory of twisted WSe$_2$ or related moir\'e materials; rather, these systems provide useful context for the field-theoretic question addressed here: how spinon pairing and Higgs fluctuations affect candidate Dirac spin liquids and their proximate paired descendants.

Spinon pairing naturally bridges candidate $\U(1)$ Dirac and proximate $\mathbb{Z}_2$ spin liquids~\cite{SSkagome,Wen2002,Lu16,Lu_2011,Lu_2017}. Here, ``Higgs criticality'' denotes the critical theory governing the fluctuations and condensation of this pairing field. In the fermionic parton approach, spin-exchange interactions decouple into hopping and singlet-pairing channels: a hopping-only Ansatz retains a $\U(1)$ invariant gauge group, whereas a finite charge-$2$ pairing amplitude reduces this to $\mathbb{Z}_2$, so microscopic interactions can energetically favor a paired state. Because the pairing field is itself gauge-charged, this variational preference is fundamentally distinct from a relevant gauge-invariant perturbation of the QED$_3$ conformal fixed point. In the continuum theory, the pairing amplitude becomes a charge-$2$ complex Higgs field $\Phi$; its condensation via the Anderson-Higgs mechanism gaps the emergent $\U(1)$ photon and yields the paired $\mathbb{Z}_2$ spin liquid.

Fermionic spinon pairing in $\mathbb{Z}_2$ spin liquids is also naturally connected to superconductivity~\cite{TSSSMV03,TSMVSS04,IvanovSenthil}, offering a possible link to superconductivity in experimental spin-liquid candidates. The triangular-lattice exact diagonalization study of Wietek {\it et al.\/}~\cite{Wietek24} shows good consistency with a $\U(1)$ spin liquid for much of the spectrum, but also finds low-energy states similar to the quantum dimer model, suggesting a crossover to a gapped $\mathbb{Z}_2$ spin liquid~\cite{SSkagome}. On the kagome lattice, the machine-learning study of Duri{\'c} {\it et al.}~\cite{Sengupta25} shows spinon pairing characteristic of a $\mathbb{Z}_2$ spin liquid.

At first sight, these results suggest a rather direct route from a gapless $\U(1)$ spin liquid to a gapped paired descendant. However, once pairing fluctuations are promoted to a dynamical charge-$2$ Higgs field in the continuum theory, a more delicate question arises: can the resulting Higgs critical theory remain well behaved once its symmetry-allowed couplings to the Dirac fermions are taken into account? The issue is, therefore, not merely whether a pairing channel exists, but whether the associated critical theory can support an extended regime of scaling, and how strongly its fate depends on the underlying lattice realization.

It is this question that we address here. In doing so, we separate the Higgs/Yukawa problem from the distinct nonperturbative question of monopoles in compact QED$_3$~\cite{Song-2019}. In a microscopic spin system the emergent $\U(1)$ gauge field is compact, and monopole operators, which insert $2\pi$ gauge flux, may be allowed by the microscopic symmetries. If such monopoles are relevant, their proliferation destroys the emergent $\U(1)_{\rm top}$ flux conservation, confines the parent $\U(1)$ Dirac spin liquid, and can preempt the Higgs critical regime studied below. Conversely, if monopoles are forbidden by symmetry, irrelevant at the scales of interest, or sufficiently suppressed over an intermediate crossover window, the noncompact QED$_3$-Higgs theory describes the scaling regime associated with condensing the charge-$2$ spinon-pairing field. Our results should, therefore, be understood as a conditional analysis of this Higgs critical regime and of its instability to symmetry-allowed Yukawa couplings, not as a complete stability analysis of compact $\U(1)$ spin liquids. This monopole problem is complementary to the one addressed here; for the maple-leaf Dirac spin liquid with $N_f=12$, it has been analyzed separately in Ref.~\cite{zhang2026}, whereas the triangular and kagome lattices were studied in Refs.~\cite{Song-2019,Song-2020}.

We study the triangular lattice and two of its depleted Archimedean tilings, the kagome and maple-leaf lattices. In each case, we begin from a candidate $\U(1)$ Dirac spin liquid and identify the symmetry-allowed spinon-pairing channel that connects it to a proximate $\mathbb{Z}_2$ spin liquid. On the triangular and kagome lattices, this $\mathbb{Z}_2$ spin liquid is the same as that found in the theory with bosonic partons~\cite{SSkagome,Lu16,Lu_2017}. 
We analyze a continuum QED$_3$-Higgs theory in which Dirac fermions couple to the emergent gauge field and to a complex charge-$2$ Higgs field. The triangular and kagome lattices give rise to the same low-energy theory with $N_f=4$ Dirac fermions, whereas the maple-leaf lattice yields an analogous theory with a larger fermion flavor number, $N_f=12$. The maple-leaf case is also distinctive in a second respect: unlike the triangular and kagome Ans\"atze, whose Dirac nodes are pinned at high-symmetry momenta, its Dirac cones can move continuously along symmetry-related lines in momentum space. These three cases, therefore, provide a controlled setting in which the general structure of the continuum theory is held fixed, while the fermionic flavor content is varied. In this sense, the comparison is not simply between three examples, but between three closely related realizations of the same broader Higgs problem.

Our analysis is based on the large-$N_{f,b}$ framework introduced by Kaul and Sachdev for relativistic gauge theories containing both fermionic and bosonic matter~\cite{Kaul08}. Within this approach, we compute the leading scaling dimensions of the Higgs field and selected fermion bilinears, together with the correlation-length exponent, and we examine in particular the symmetry-allowed Yukawa coupling between the Higgs field and the Dirac fermions. This coupling is central to the fate of the would-be Higgs critical point. We find that fluctuations of the Higgs field reduce the relevance of this coupling compared to the $N_b=0$ case. Although this suppression is insufficient to drive the coupling to irrelevance at leading order---meaning the fixed point remains asymptotically unstable at the longest scales on all three lattices---the details of this instability differ. Because the triangular and kagome lattices share the same continuum theory, they receive identical leading corrections. By contrast, the maple-leaf lattice presents a competing scenario: its larger value of $N_f$ suppresses the primary Yukawa instability much more strongly. The maple-leaf case is, therefore, of particular interest, since the larger flavor number suppresses this instability most strongly and may allow for a broader intermediate regime governed by Higgs critical behavior. Indeed, our results suggest that models with even higher fermion numbers---particularly those where the Dirac nodes remain pinned at high-symmetry points---would be worth investigating as potential candidates for a fully stable Higgs fixed point.

These results place Higgs transitions out of nonbipartite $\U(1)$ Dirac spin liquids within the broader discussion of fermionic deconfined criticality~\cite{Senthil_2004,Senthil2004a,Christos_2024,Christos_2023,Feuerpfeil_2026,Maity-2026}. More generally, they bear on a question that has become increasingly important in modern studies of fermionic criticality: when a nominal fixed point is destabilized by a symmetry-allowed perturbation, to what extent can it nevertheless govern observable intermediate-scale physics? In that broader setting, the significance of the present work is not that it establishes a stable asymptotic fixed point for these lattices, but that it clarifies how lattice-dependent low-energy structure enters a common continuum Higgs problem to shape the extent and character of the scaling regime. From this point of view, the triangular, kagome, and maple-leaf lattices offer a powerful comparative setting for examining how microscopic realization leaves its imprint on fermionic criticality in frustrated quantum magnets.

\section{U(1) continuum theories and Higgs transitions on depleted triangular lattices}\label{sec:higgs_theories}
We emphasize at the outset that our analysis does not start from a single microscopic spin-exchange Hamiltonian whose phase diagram is derived in this work. Instead, we take symmetry-allowed fermionic parton {\it Ansätze} for candidate $\U(1)$ Dirac spin liquids---motivated by previous numerical, variational, and projective symmetry group (PSG) studies on the triangular lattice and its depleted relatives, the kagome and maple-leaf geometries~\cite{Iqbal-2011,Iqbal-2013,Iqbal-2015,Iqbal-2016,Iqbal-2016_breathing,Mei-2017,He-2017,Hu-2019,Ferrari-2019,Drescher-2025,Jiang-2026,Kovalska-2026}---as parent states, and ask which symmetry-allowed spinon-pairing perturbations connect them to proximate $\mathbb{Z}_2$ spin liquids, and what continuum QED$_3$-Higgs theories result. The microscopic spin models cited here, therefore, serve only to motivate the candidate low-energy parton states; the field-theoretic analysis itself focuses on the Higgs transition out of those states, deriving the low-energy Dirac Lagrangian, identifying the symmetry-allowed pairing terms corresponding to the various $\mathbb{Z}_2$ {\it Ansätze}, and calculating their resulting scaling dimensions.

To formulate the fermionic spinon theory, we begin by expressing the spin operators with respect to spinons $f_{\vi\alpha}, \alpha=\uparrow, \downarrow$ for sites $\vi=(i_x,i_y)$~\cite{Abrikosov-1965}:
\begin{equation}
    \mbf{S}_{\vi}=\frac{1}{2}f_{\vi\alpha}^\dag \bm{\sigma}_{\alpha\beta}f_{\vi\beta}\,.
\end{equation}
Following Wen~\cite{Wen2002}, we introduce a Nambu spinor 
\begin{equation}
    \Psi_{\vi}=\begin{pmatrix}
        f_{\vi\uparrow}\\f_{\vi\downarrow}^\dag
    \end{pmatrix}
\end{equation}
and a mean-field link field $u_{\vi\vj}$, which leads to the Bogoliubov Hamiltonian
\begin{equation}\label{eq:bogoliubov_hamiltonian}
    H_\mathrm{MF}=-\sum_{\vi\vj}\Psi_{\vi}^\dag u_{\vi\vj}\Psi_{\vj}\,.
\end{equation}
Here, 
\begin{equation}
    u_{\vi \vj} = iu_{\vi\vj}^0\tau^0+u_{\vi\vj}^a\tau^a\,,
\end{equation}
with the Pauli matrices $\tau^\mu$ acting on the Nambu spinor $\Psi_{\vi}$. Spin-rotation symmetry implies that $u_{\vi\vj}^\mu \in \mathbb{R}$ and 
\begin{equation}
    u_{\vj\vi}^0=-u_{\vi\vj}^0, \quad  u_{\vj\vi}^a=u_{\vi\vj}^a\,.
\end{equation}
The spinon representation is invariant under $\SU(2)_g$ gauge transformations~\cite{Affleck1988,Dagotto-1988} for a symmetry operator $g$:
\begin{equation}
\begin{split}
    \SU(2)_g: \Psi_{\vi}&\rightarrow W_{g,g(\vi)}\Psi_{g(\vi)}\,,\\
    u_{\vi\vj}&\rightarrow W_{g,g(\vi)}u_{g(\vi),g(\vj)}W^\dag_{g,g(\vj)}\,.
\end{split}    
\end{equation}

Our analysis of the three studied lattices---triangular, kagome, and maple-leaf---proceeds as follows. We use the Nambu formalism above to define the PSG and the microscopic link matrices. Fourier transforming these link matrices gives the momentum-space mean-field Hamiltonian $H_{\rm MF}(\vk)$, i.e., the spinon tight-binding (Bloch) Hamiltonian, whose diagonalization yields the spinon dispersion. The Dirac points $\vK_v$ are the momenta where two spinon bands become degenerate, and the corresponding degenerate eigenvectors $\phi_m^v$ of $H_{\rm MF}(\vK_v)$ furnish a natural basis for the low-energy subspace. This step is not merely technical: it is the continuum Dirac description, not the microscopic lattice details, that fixes the universality class of the eventual Higgs transition, so very different lattices can flow to the same low-energy QED$_3$ theory whenever they share the same number and symmetry structure of Dirac cones~\cite{Feuerpfeil_2026,Maity-2026}. Expanding $f_{\vk\sigma}$ near each $\vK_v$ in the basis $\phi_m^v$ defines the low-energy Dirac fields $\psi_{\vq,v,\sigma,m}$ via
\begin{equation}\label{eq:low_energy_modes}
    f_{\vK_v+\vq,\sigma}\approx \sum_{m}\phi_m^v\,\psi_{\vq,v,\sigma,m}\,,
\end{equation}
with $\vq$ measured relative to $\vK_v$. Here, $\rho^a$ act on the two-component Dirac subspace $m$, $\mu^a$ on the valley subspace $v$, and $\sigma^a$ on the physical spin $\sigma$.

We then identify which additional pairing terms $\delta u_{\vi\vj}$ are allowed by the microscopic symmetries, implemented projectively through the PSG of the descendant $\mathbb{Z}_2$ \textit{Ansatz}; these break the $\U(1)$ invariant gauge group of the hopping-only parent state down to $\mathbb{Z}_2$. Expanding the perturbed mean-field Hamiltonian to lowest order in the low-energy modes, the allowed perturbations appear as prefactors of fermion bilinears, which we promote to a complex, charge-$2$ Higgs field $\Phi$: condensing $\Phi$ yields the paired $\mathbb{Z}_2$ descendant of the parent $\U(1)$ Dirac spin liquid. At criticality $\langle \Phi\rangle=0$ and the Higgs field is massless, requiring us to study its fluctuations at the level of the QFT. This construction also fixes the flavor counting: the triangular and kagome states host two Dirac cones and two spin components, giving $N_f = 2\times2 = 4$, while the maple-leaf state hosts six Dirac cones, giving $N_f = 6\times2 = 12$. The \textit{Ansätze}, Dirac cones, and Yukawa couplings for each lattice are summarized in \tabref{tab:summary}.

\subsection{Triangular lattice}
The microscopic motivation for the triangular lattice comes primarily from the spin-$\frac{1}{2}$ $J_1$--$J_2$ Heisenberg antiferromagnet, where several DMRG, exact-diagonalization, variational Monte Carlo, and dynamical studies report an intermediate spin-liquid regime consistent with a $\U(1)$ Dirac spin liquid~\cite{Zhu-2015,Hu-2015,Iqbal-2016,Hu-2019,Ferrari-2019,Wietek24,Sherman-2023,Drescher-2025,Jiang-2026,Kovalska-2026}. We use the corresponding $\pi$-flux $\U(1)$ parton {\it Ansatz} (\#20 in Ref.~\cite{Lu16}) as the parent state, whose proximate symmetric $\mathbb{Z}_2$ descendant is the spinon-pairing state classified in Ref.~\cite{Lu16}.

The details of the {\it Ansatz} and its PSG are provided in \appref{app:ansatz_triangular}. To accommodate the $\pi$-flux and maintain translational invariance, we double the unit cell in the $y$-direction. The resulting $\mathrm{U}(1)$ mean-field Hamiltonian \eqnref{eq:bogoliubov_hamiltonian} can be expressed in momentum space as
\begin{widetext}
\begin{equation}
\begin{split}\label{eq:H_MF_tri}
    H_\mathrm{MF} &= -t \sum_{\vk,\sigma}f_{\vk\sigma}^\dag 
    \begin{pmatrix}
        -2\cos(k_x) & 1-e^{ik_x}+e^{i(k_x-\sqrt{3}k_y)}+e^{-i\sqrt{3}k_y}\\
         1-e^{-ik_x}+e^{-i(k_x-\sqrt{3}k_y)}+e^{i\sqrt{3}k_y} & 2\cos(k_x)
    \end{pmatrix}f_{\vk\sigma}\,.
\end{split}
\end{equation}
\end{widetext}
With reciprocal lattice vectors $\vb_1=2\pi(1,1/\sqrt{3})$ and $\vb_2=2\pi(0,1/\sqrt{3})$, this Hamiltonian features two Dirac points at $\vK_v=\pm(\pi/2,-\pi/(2\sqrt{3}))$, as shown in \appref{app:ansatz_triangular}. We associate these with a valley degree of freedom $v=\pm$, described by Pauli matrices $\mu^a$.

To derive the low-energy theory, we expand around the Dirac points using the following basis:
\begin{equation}
\begin{split}
    \phi_1^+&=\frac{1}{\sqrt{6+2\sqrt{3}}}(1-i,1+\sqrt{3})^\top\,,\\ 
    \phi_2^+&=\frac{1}{\sqrt{6-2\sqrt{3}}}(1-i,1-\sqrt{3})^\top\,,\\
    \phi_1^-&=\frac{1}{\sqrt{6-2\sqrt{3}}}(1+i,1-\sqrt{3})^\top\,,\\
    \phi_2^-&=\frac{-1}{\sqrt{6+2\sqrt{3}}}(1+i,1+\sqrt{3})^\top\,.\\
\end{split}
\end{equation}
The low-energy modes $\psi_{\vq,v,\sigma,m}$ are defined as in \eqnref{eq:low_energy_modes}, where $m$ is the Dirac index associated with Pauli matrices $\rho^a$. This leads to the linearized Dirac Hamiltonian:
\begin{equation}
\begin{split}
    H_\mathrm{MF}&= -\sqrt{6}t\sum_{\vq}\psi_{\vq}^\dag \left(\rho^xq_x+\rho^yq_y\right) \psi_{\vq}\,.
\end{split}
\end{equation}

Defining gamma matrices $\gamma^0 =\rho^z$, $\gamma^x = -i\rho^y$, $\gamma^y = i\rho^x$, and the adjoint $\bar{\psi}=\psi^\dag\gamma^0$, we arrive at the relativistic Dirac Lagrangian:
\begin{equation}
\begin{split}\label{eq:triangular_MF}
    \mathcal{L}_\mathrm{MF}&=i\bar{\psi}_{v\sigma}\gamma^\mu(\partial_\mu-ia_\mu)\psi_{v\sigma}\,,
\end{split}
\end{equation}
where we have adopted the Lorentz signature $(1,-1,-1)$ and absorbed the hopping $t$ into the speed of light.

Next, we consider the pairing term that gaps the Dirac nodes and reduces the invariant gauge group to $\mathbb{Z}_2$:

\begin{equation}
\begin{split}\label{eq:higgs_dirac}
    \delta H &= -\lambda \sum_{\vecr}\left(f_{\vecr\uparrow}^\dag f_{\vecr\downarrow}^\dag+\text{h.c.}\right) \\
    &\approx -\lambda \int \dd^2 r \psi^\top (i\mu^yi\rho^yi\sigma^y)\psi+\text{h.c.}
\end{split}
\end{equation}
This term corresponds to zero-momentum pairing and therefore couples only valleys at opposite momenta.

In order to capture fluctuations around $\langle\lambda\rangle=0$ in the $\U(1)$ Dirac spin liquid, we will promote $\lambda$ to a dynamic, complex Higgs field $\Phi$. This field couples to the fermions via the Yukawa term
\begin{equation}
    \mathcal{L}_{\Phi\psi}= y\Phi^* \psi^{\top}(i\mu^y\gamma^x i\sigma^y)\psi+\text{h.c.}
\end{equation}
and has a gauge charge of $2$. Condensation of the Higgs field, $\langle \Phi\rangle \propto \lambda$, drives the transition to the gapped $\mathbb{Z}_2$ spin liquid, where $\lambda$ corresponds to the singlet superconducting pairing $\Delta_{\mathrm{ssc}}$ identified in Ref.~\cite{Lu16}. 

The resulting quantum field theory is governed by the Lagrangian
\begin{equation}
\begin{split}\label{eq:triangular_QFT}
    \mathcal{L} &= i\bar{\psi}_{v\sigma}\gamma^\mu(\partial_\mu-ia_\mu)\psi_{v\sigma} + \frac{1}{2g}|(\partial_\mu-2ia_\mu)\Phi|^2
\end{split}
\end{equation}
and the Yukawa coupling $\mathcal{L}_{\Phi\psi}$. This theory corresponds to $\mathrm{QED}_3$ with $N_f=4$ fermion flavors ($2$ valleys $\times$ $2$ spins) coupled to a complex charge-$2$ Higgs field.

\subsection{Kagome lattice}
For the kagome lattice, the relevant microscopic setting is the spin-$\frac{1}{2}$ kagome Heisenberg antiferromagnet and its extensions~\cite{Iqbal-2016_breathing,Iqbal-2021,Iqbal-2011_vbc,Iqbal_2012}. The $\U(1)$ Dirac spin liquid has long been identified as a highly competitive projected-fermion variational state for the nearest-neighbor model~\cite{Ran-2007,Iqbal-2013}, and numerical studies have found low-energy signatures consistent with Dirac cones in the spectrum~\cite{He-2017}. Proximate $\mathbb{Z}_2$ spin liquids near this $\U(1)$ parent state have been classified within the fermionic PSG framework~\cite{Lu_2011,Lu_2017} and studied variationally~\cite{Iqbal-2011,Mei-2017,Iqbal-2014_gap}. As promising candidate spin liquids, we, therefore, take the $[0,\pi]$ $\U(1)$ state and its gapped $\mathbb{Z}_2[0,\pi]\beta$ descendant in our comparative Higgs analysis.

To maintain translational invariance, we enlarge the unit cell in the $x$-direction. This introduces a sublattice index $\{A,B\}$ in addition to the three-site kagome basis $s \in \{1,2,3\}$, as detailed in \appref{app:kagome}. Ordering the six resulting basis states as $(1A, 2A, 3A, 1B, 2B, 3B)$, we express the mean-field Hamiltonian in momentum space as
\begin{widetext}
\begin{equation}
    H_\mathrm{MF} = -\sum_{\vk,\sigma}f_{\vk\sigma}^\dag
    \begin{pmatrix}
        H_{AA}(P,Q) & H_{AB}(P,Q) \\
        H_{AB}^\dag(P,Q) & H_{AA}(P,-Q)
    \end{pmatrix} 
    f_{\vk\sigma}\,,
\end{equation}
where $P \equiv e^{2ik_x}$ and $Q \equiv e^{i(k_x/2+\sqrt{3}k_y/2)}$. The block matrices are given by:
\begin{equation}
    H_{AA} = \begin{pmatrix}
        \lambda_z & t_1(1+Q) & t_1+t_2 Q \\
        t_1(1+Q^{-1}) & \lambda_z & t_1+t_2 Q^{-1} \\
        t_1+t_2 Q^{-1} & t_1+t_2 Q & \lambda_z 
    \end{pmatrix}\,,\quad
    H_{AB} = \begin{pmatrix}
        0 & t_2(1-P^{-1} Q ) & t_1+t_2 Q^{-1} \\
        t_2(P^{-1}+Q^{-1}) & 0 & t_2+t_1 Q^{-1} \\
        P^{-1}(t_1-t_2 Q) & P^{-1}(t_2-t_1 Q) & 0 
    \end{pmatrix}\,.
\end{equation}
\end{widetext}
The reciprocal lattice vectors are $\vb_1=\pi(1,-1/\sqrt{3})$ and $\vb_2=4\pi(0,1/\sqrt{3})$. Enforcing the half-filling constraint requires tuning the onsite potential to
\begin{equation}
\lambda_z = (\sqrt{3}-1)t_1+(1+\sqrt{3})t_2\,,
\end{equation}
at which point the spectrum develops two Dirac nodes at $\vK_v=\pm(\frac{\pi}{2},\frac{\pi}{2\sqrt{3}})$. As before, we associate these with a valley index $v=\pm$.

Following the convention in Ref.~\cite{Hermele2008}, we choose the following basis for the eigenvectors at the Dirac points:
\begin{equation}
\begin{split}
    \phi_1^+&=\frac{1}{\sqrt{6}}(-i,e^{i2\pi/3},\sqrt{2}e^{i\pi/12},i,e^{i\pi/6},0)^\top\,,\\
    \phi_2^+&=\frac{1}{\sqrt{6}}(e^{-i5\pi/12},e^{-i\pi/12},0,e^{-i5\pi/12},e^{i5\pi/12},-\sqrt{2})^\top\,,\\
    \phi_1^-&=\frac{1}{\sqrt{6}}(e^{i5\pi/12},e^{i\pi/12},0,e^{i5\pi/12},e^{-i5\pi/12},-\sqrt{2})^\top\,,\\
    \phi_2^-&=\frac{-1}{\sqrt{6}}(i,e^{-i2\pi/3},\sqrt{2}e^{-i\pi/12},-i,e^{-i\pi/6},0)^\top\,.\\
\end{split}
\end{equation}
Expanding the Hamiltonian in terms of the low-energy modes \eqnref{eq:low_energy_modes} yields the Dirac form:
\begin{equation}
    H_{\mathrm{MF}}\propto -\sum_{\vq}\psi_{\vq}^\dag (\rho^xq_x+\rho^yq_y)\psi_{\vq}\,.
\end{equation}
Note that a perturbation in the second-neighbor hopping $t_2 \rightarrow t_2 + \delta t_2$ to first order shifts the chemical potential, $\delta H \propto (1+\sqrt{3})\delta t_2 \psi_{\vq}^\dag \psi_{\vq}$, which can be re-absorbed into the definition of $\lambda_z$.

To analyze the pairing instabilities, we note that, in the PSG/Nambu notation of Appendix~\ref{app:kagome}, the pairing components $\Delta_1, \Delta_2,$ and $\lambda_x$ are obtained from the corresponding hopping and onsite components $t_1, t_2,$ and $\lambda_z$ by replacing the associated $\tau^z$ link matrices by $\tau^x$ link matrices. Projected into the spinon low-energy basis used here, the resulting leading-order perturbation at $\vK_v$ is thus:
\begin{equation}
\begin{split}\label{eq:kagome_z2_perturbation_lattice}
\delta H &\approx  -\Delta \int \dd^2 r \psi^\top (i\mu^yi\rho^yi\sigma^y)\psi+\text{h.c.}\,,
\end{split}
\end{equation}
where $\Delta = -(\sqrt{3}-1)\Delta_1 - (1+\sqrt{3})\Delta_2 + \lambda_x$. This matches the structure of the triangular lattice pairing in \eqnref{eq:higgs_dirac}. We can, therefore, introduce a dynamical Higgs field $\Phi$ such that condensation $\langle \Phi \rangle \propto \Delta$ breaks the gauge symmetry to $\mathbb{Z}_2$. As established in Ref.~\cite{Lu_2011}, this is the unique symmetry-allowed mass term that opens a gap. Consequently, the kagome lattice flows to the same continuum Lagrangian \eqnref{eq:triangular_QFT} as the triangular lattice, characterized by $N_f=4$ fermion flavors.

The observation that both the triangular and kagome lattices are governed by the same QFT can be conceptually understood in the following way: Both $\U(1)$ mean-field \textit{Ansätze} host two Dirac cones at high-symmetry points, leading to the same QED$_3$ theory with $N_f=4$ Dirac fermions. As the Dirac cones are pinned, the zero-momentum pairing terms gap out the Dirac cones on both lattices. Even though the realization of these pairing terms on the parton-level differs between the two cases, after projection to the Dirac fermions, condensing the pairing terms leads to the same fermion mass term $\Phi^*\psi^\top M \psi$ with $M=i\mu^y\gamma^xi\sigma^y$.

\subsection{Maple-leaf lattice}

For the maple-leaf lattice, recent numerical and variational studies of spin-$\frac{1}{2}$ Heisenberg-type models have reported extended quantum-disordered or candidate spin-liquid regimes~\cite{Gresista2023,Beck-2024,Ebert2026a,Ebert2026b,Gresista2026,Schmoll2026}. These results motivate exploring fermionic parton \textit{Ansätze} on the maple-leaf geometry. Rather than restricting our focus to a specific microscopic Hamiltonian that exactly stabilizes the uniform $\U(1)$ state, we treat this translationally invariant $\U(1)$ \textit{Ansatz} as a symmetry-allowed, large-flavor Dirac spin-liquid candidate~\cite{zhang2026}. This allows us to systematically analyze the Higgs theory governing its proximate $\mathbb{Z}_2$ descendant.

Unlike the triangular and kagome cases, the $\mathrm{U}(1)$ {\it Ansatz}, detailed in \appref{app:mapleleaf}, on the maple-leaf lattice is translationally invariant without the need for unit cell doubling~\cite{Sonnenschein-2024}. This allows us to construct the momentum-space Hamiltonian directly. Defining the phase factors $P=e^{i(3\sqrt{3}k_x-k_y)/2}$ and $Q=e^{i(5k_y-\sqrt{3}k_x)/2}$, we obtain the $6 \times 6$ mean-field Hamiltonian:
\begin{widetext}
\begin{equation}
\begin{split}
    H_\mathrm{MF}=-\sum_{\vk,\sigma}f_{\vk\sigma}^\dag
    \begin{pmatrix}
        \lambda_z & t_h & t_t PQ & t_d Q & t_t Q & t_h  \\
        t_h & \lambda_z & t_h & t_t Q & t_d P^{-1} & t_t P^{-1}  \\
        t_t P^{-1}Q^{-1} & t_h & \lambda_z & t_h & t_t P^{-1} & t_d P^{-1}Q^{-1} \\
        t_d Q^{-1} & t_t Q^{-1} & t_h & \lambda_z & t_h & t_t P^{-1}Q^{-1} \\
        t_t Q^{-1} & t_d P & t_t P & t_h & \lambda_z & t_h  \\
        t_h & t_t P & t_d PQ & t_t PQ & t_h & \lambda_z \\
    \end{pmatrix}
    f_{\vk\sigma}\,.
\end{split}
\end{equation}
\end{widetext}
\begin{table*}[th!]
    \centering
    \begin{tblr}{
          width = 0.9\linewidth,
          colspec = {c | X[c] X[c] X[c] },
          rowsep = 3pt,
          cells = {valign=m},
          cells = {font=\linespread{1.3}\selectfont}
        }\hline\hline
          & Triangular &Kagome & Maple-leaf \\\hline 
         Reciprocal lattice vectors & $\vb_1=2\pi(1,1/\sqrt{3})$ $\vb_2=2\pi(0,1/\sqrt{3})$ & $\vb_1=\pi(1,-1/\sqrt{3})$ $\vb_2=4\pi(0,1/\sqrt{3})$ & $\vb_1=2\pi/21(5\sqrt{3},3)$ $\vb_2=2\pi/21(\sqrt{3},9)$\\\hline
         $\U(1)$ Ansatz & $\U(1)$ $\pi$-flux~\cite{Lu16}  &$[0,\pi]$ $\U(1)$~\cite{Lu_2011} & UC00~\cite{Sonnenschein-2024}\\
        Dirac cones & $\pm(\pi/2,-\pi/(2\sqrt{3}))$& $\pm (\pi/2,\pi/(2\sqrt{3}))$& $C_6^v \vK$, $\vK\in \overline{\Gamma\mathrm{M}}$, $v=1,\dots,6$ \\
        $N_f$ & $4$& $4$& $12$\\ \hline
        $\Z_2$ Ansatz & \#20 in Ref.~\cite{Lu16} & $\mathbb{Z}_2[0,\pi]\beta$~\cite{Lu_2011}& Z0002~\cite{Sonnenschein-2024}\\ 
        Yukawa coupling & $y\Phi^*\psi^\top(i\mu^y\gamma^xi\sigma^y)\psi+\text{h.c.}$&$y\Phi^*\psi^\top(i\mu^y\gamma^xi\sigma^y)\psi+\text{h.c.}$ & $y \Phi^* \psi_v^\top (\gamma^y\sigma^y)\psi_{-v}+y_2 \Phi^* \psi_v^\top (i\sigma^y)\psi_{-v}+ \text{h.c.}$\\
        $N_b$ & $1$& $1$& $1$\\\hline
    \end{tblr}
    \caption{Overview of the parton mean-field constructions on the triangular, kagome, and maple-leaf lattices. For each lattice we list the reciprocal lattice vectors, the parent $\U(1)$ Ansatz, and the momentum-space location of the Dirac cones in the spinon dispersion, together with the resulting number of Dirac fermion flavors $N_f$. We then give the $\Z_2$ Ansatz obtained as a pairing perturbation of the parent state, which reduces the invariant gauge group from $\U(1)$ to $\Z_2$, and report the continuum Yukawa coupling that this perturbation generates between the Higgs field $\Phi$ and the low-energy Dirac spinons, along with the number of complex Higgs-field flavors $N_b$.}
    \label{tab:summary}
\end{table*}
The reciprocal lattice vectors are $\vb_1=2\pi/21(5\sqrt{3},3)$ and $\vb_2=2\pi/21(\sqrt{3},9)$. We focus on a representative parameter regime around $t_t=0$ and $t_h=-0.75t_d$, where the spectrum features six Dirac points along the $\overline{\Gamma \mathrm{M}}$ lines, related by $C_6$ symmetry.

These Dirac nodes are inherently anisotropic. To map the low-energy physics to a standard Dirac form, we choose a local basis $\phi_m^v$ for each valley $v=1,\dots,6$ and introduce rotated and rescaled coordinates $(\tilde{q}_x^v, \tilde{q}_y^v)$ to restore isotropy. In this frame, the mean-field Hamiltonian takes the universal form:
\begin{equation}
    H_{\mathrm{MF}}\propto -\sum_{\tilde{\vq}}\psi_{\tilde{\vq}}^\dag(\rho^x\tilde{q}_x+\rho^y\tilde{q}_y)\psi_{\tilde{\vq}}\,.
\end{equation}

For the purposes of the continuum theory, we neglect the weak velocity anisotropy (the ratio of principal velocities is approximately $1.19$ for the chosen parameters).

Expanding the microscopic pairing and hopping perturbations in the low-energy modes yields:

\begin{equation}
\begin{split}\label{eq:maple_Z2_perturbations}
    \delta H \propto \sum_{v}& (\Delta_h+\Delta_d)\psi_v^\top(i\sigma^y)\psi_{-v}\\
    &-(\lambda_x+\Delta_t)\psi_v^\top (i\sigma^y\rho^x)\psi_{-v}+\text{h.c.}
\end{split}
\end{equation}
The second term contains a Dirac-index matrix, $\rho^x=-i\gamma^y$, and is the maple-leaf counterpart of the Dirac-matrix Yukawa coupling appearing on the triangular and kagome lattices. Its physical effect is different, however. In PSG language, the Z0002 descendant is obtained from the UC00 state by adding $\tau^x$ components in the same translation- and $C_6$-symmetric bond and onsite channels as the $\tau^z$ hopping parameters. The PSG, therefore, fixes the low-energy form factor of this perturbation: after projection to the two-band Dirac subspace it appears as the same Dirac matrix that multiplies one component of the kinetic momentum, rather than as a matrix that anticommutes with the full Dirac Hamiltonian.

Equivalently, for a pair of opposite valleys, the kinetic Hamiltonian may be written schematically as $H_0=\tau^z(\rho^x\tilde q_x+\rho^y\tilde q_y)$, where $\tau^a$ acts on the Nambu space. The Dirac-matrix Higgs perturbation has the structure $m\,\tau^x\rho^x$ up to a basis convention. Since $\tau^x\rho^x$ anticommutes with $\tau^z\rho^x$ but commutes with $\tau^z\rho^y$, it is not a Dirac mass. Rather, it has the form of a valley-dependent vector-potential perturbation, with spectrum $E^2=\tilde q_x^2+(\tilde q_y\pm m)^2$ up to a relabeling of the local axes. It, therefore, shifts the nodal positions along the symmetry-related $\overline{\Gamma\mathrm{M}}$ lines instead of lifting the nodes. This is possible because the maple-leaf Dirac points sit at generic points on these lines: $C_6$ symmetry relates the six nodes to one another but does not pin each node to a time-reversal-invariant or other high-symmetry momentum. A conventional mass would have to anticommute with both kinetic matrices; this is the structure generated by the pinned triangular and kagome pairing perturbations, but not by the maple-leaf Dirac-matrix channel. The first term in \eqnref{eq:maple_Z2_perturbations} gives the additional Dirac-trivial maple-leaf coupling, denoted $y_2$ below, and is kept as a separate allowed perturbation.

This results in the same emergent $\mathrm{QED}_3$-Higgs theory \eqnref{eq:triangular_QFT} with $N_f=12$ fermion flavors, but an altered Yukawa coupling:

\begin{equation}
\begin{split}
    \mathcal{L}_{\Phi\psi,\mathrm{ML}} &= y \sum_v \Phi^* \psi_v^\top (\gamma^y\sigma^y)\psi_{-v}\\
    &+y_2 \sum_v \Phi^* \psi_v^\top (i\sigma^y)\psi_{-v}+ \text{h.c.}
\end{split}
\end{equation}
This flavor-rich continuum description serves as the basis for our comparative RG analysis, where the increased $N_f$ is expected to further suppress the relevance of the Yukawa coupling.

\section{Renormalization group study of the U(1) Dirac spin liquids}
As discussed in \secref{sec:higgs_theories}, the continuum limits for all three lattices are governed by QED$_3$ with $N_f=4$ ($12$ for the maple-leaf) complex Dirac fermions and $N_b=1$ complex Higgs fields with a gauge charge of $2$. This emergent unification---which is unattainable at the microscopic level due to distinct spin configurations---parallels the unified description of the $\SU(2)$ $\pi$-flux state across the square, Shastry-Sutherland, and checkerboard lattices~\cite{Feuerpfeil_2026,Maity-2026}.

We now investigate the effect of these matter fields on the critical exponents and stability of the $\U(1)$ Dirac spin liquids. To this end, we perform a controlled large-$N_{f,b}$ expansion. We introduce a flavor index $\alpha=1,\dots,N_b$ for the Higgs field and generalize the fermion valley index to $v=1,\dots,N_f/2$, requiring $N_f$ to be a multiple of $4$ to ensure every valley $v$ maintains an inversion-symmetric partner $-v$. We define $\mu^y$ to act as the Pauli matrix on the space $(v,-v)$, and vanish for any other pair of valleys. This yields the continuum Lagrangian with a non-Lorentz-invariant Yukawa coupling:
\begin{equation}
\begin{split}\label{eq:ml_yukawa}
    \mathcal{L}&=i\bar{\psi}_{v\sigma}\gamma^\mu(\partial_\mu-ia_\mu)\psi_{v\sigma}+\frac{1}{2g}|(\partial_\mu-2ia_\mu)\Phi_{\alpha}|^2\,,\\
    \mathcal{L}_{\Phi\psi}&=y\sum_{\alpha}\Phi_{\alpha}^*\psi^\top(i\mu^y\gamma^xi\sigma^y)\psi+\text{h.c.}\,,
\end{split}
\end{equation}
as well as 
\begin{equation}
\begin{split}\label{eq:yukawa_ml}
    \mathcal{L}_{\Phi\psi,\mathrm{ML}} &= y \sum_{\alpha,v} \Phi_\alpha^* \psi_v^\top (\gamma^y\sigma^y)\psi_{-v}\\
    &+y_2 \sum_{\alpha,v} \Phi_\alpha^* \psi_v^\top (i\sigma^y)\psi_{-v}+ \text{h.c.}
\end{split}
\end{equation}
for the maple-leaf lattice.
Similar QED$_3$ Lagrangians, wherein a charge-$2$ scalar couples to $N_f$ Dirac fermions via a Yukawa term, were investigated in Refs.~\cite{Boyack_2018,Dumitrescu_2026} but only for Lorentz-invariant Yukawa couplings. We will use the Euclidean metric in the following.

 To resolve the constraint $\sum_{\alpha} |\Phi_{\alpha}|^2=1$, we add a real-valued field $\lambda$ as a Lagrange multiplier, which allows us to write the bosonic action by rescaling $\Phi_{\alpha} \rightarrow \sqrt{\frac{g}{N_b}}\Phi_{\alpha}$ as
\begin{equation}
    \mathcal{L}_\Phi = \frac{1}{2g}\left[|(\partial_\mu-2ia_\mu)\Phi_{\alpha}|^2+i\lambda \left(|\Phi_{\alpha}|^2-\frac{N_b}{g}\right)\right]\,.
\end{equation}

\begin{figure}
    \centering
    \includegraphics[width=0.8\linewidth]{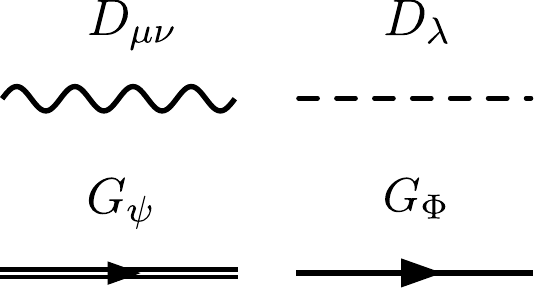}
    \caption{Definition of diagrammatic symbols for the propagators $D_{\mu\nu},D_\lambda,G_\psi,$ and $G_\Phi$.}
    \label{fig:propagator_symbols}
\end{figure}

We begin by integrating out the bosons to derive the corresponding saddle-point equation~\cite{Polyakov_1987}:
\begin{equation}
    \frac{1}{g}=\int \frac{\dd^3 k}{(2\pi)^3}\frac{1}{k^2+i\lambda}\,.
\end{equation}
This is solved by setting $r=i\lambda_0$, yielding the bare Higgs propagator $G_\Phi(k)=(k^2+r)^{-1}$. In the large-$N_b$ limit, the theory becomes critical when the mass parameter is tuned to $r=0$~\cite{Kaul08}.

\begin{figure}
    \centering
    \includegraphics[width=0.9\linewidth]{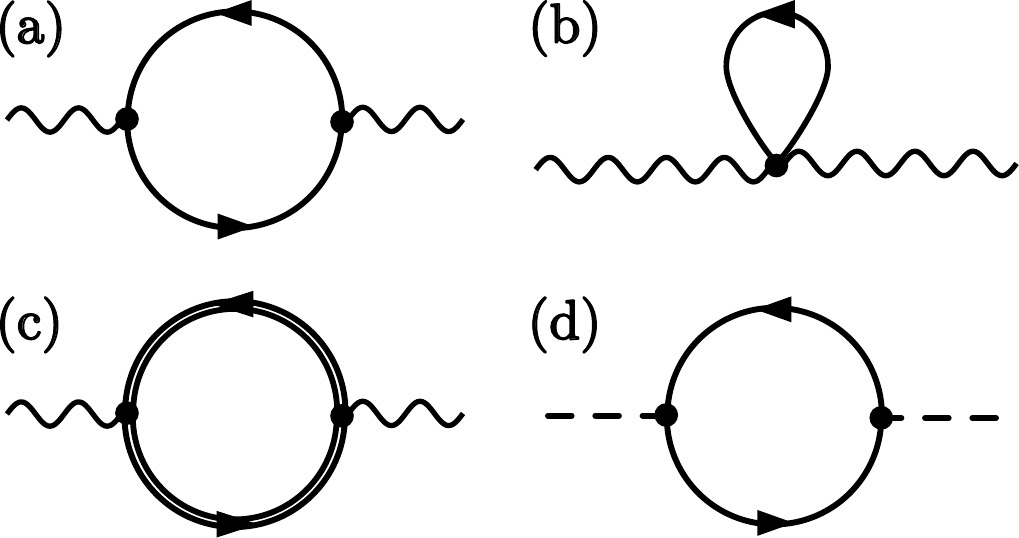}
    \caption{One-loop diagrams determining the effective action at large $N_f$ and $N_b$. Diagrams (a), (b), and (c) contribute to the vacuum polarization of the gauge field $a_\mu$, while (d) contributes to the polarization of the auxiliary field $\lambda$.}
    \label{fig:saddle_point_expansion_diagrams}
\end{figure}
The propagators for $a_\mu$, $\lambda$, $\psi$, and $\Phi$ are shown in Fig.~\ref{fig:propagator_symbols}. We obtain the effective gauge theory for the fluctuating fields $\lambda$ and $a_{\mu}$ at $T=0$ by expanding around this saddle point. The Feynman diagrams corresponding to the leading quantum corrections are shown in \figref{fig:saddle_point_expansion_diagrams}. Up to numerical prefactors, these vacuum polarization integrals are identical to those evaluated in Refs.~\cite{Christos_2024,Feuerpfeil_2026}. Integrating out the matter fields leads to the quadratic effective action:
\begin{equation}
    \mathcal{F}=\frac{1}{2}\int\frac{\dd^3 p}{(2\pi)^3}\left(\Pi_\lambda\lambda^2+\Pi_{a}\left(\delta_{\mu\nu}-\frac{p_\mu p_\nu}{p^2}\right)a_\mu a_\nu\right)\,,
\end{equation}
where the polarization functions are given by
\begin{equation}
\begin{split}\label{eq:polarization_lambda_A}
    \Pi_\lambda(p,r)&=\frac{N_b}{4\pi p}\arctan \frac{p}{2\sqrt{r}}\,,\\
    \Pi_a(p,r)&=N_f\frac{p}{16} + 4N_b\left[\frac{p^2+4r}{8\pi p}\arctan \frac{p}{2\sqrt{r}}-\frac{\sqrt{r}}{4\pi}\right]\,.
\end{split}
\end{equation}

The resulting propagators are obtained by imposing the covariant gauge $k_\mu a_\mu=1-\zeta$ parameterized by $\zeta$:
\begin{equation}
\begin{split}
    D_\lambda &=\langle \lambda \lambda \rangle = \frac{1}{\Pi_\lambda}\,,\\
    D_{\mu\nu} &=\langle a_\mu a_\nu\rangle = \frac{1}{\Pi_a}\left(\delta_{\mu\nu}-\zeta \frac{p_\mu p_\nu}{p^2}\right)\,,\\
    G_\Phi &= \frac{1}{k^2+r}\,,\\
    G_\psi &= \frac{\slashed{k}}{k^2}\,.
\end{split}
\end{equation}
Because we are interested in the critical point, we will set $r=0$ in the following analysis, which simplifies the polarizations in \eqnref{eq:polarization_lambda_A} to their critical forms:
\begin{equation}
    \Pi_\lambda(p) = \frac{N_b}{8 p}\,, \quad \Pi_a(p)=(N_f+4N_b)\frac{p}{16}\,.
\end{equation}

\subsection{Fermion self-energy}
The scaling dimension of the fermion field $\psi$ is defined as
\begin{equation}
\dim[\psi]=\frac{D-1+\eta_\psi}{2}\,,
\end{equation}
where $D=3$ and $\eta_\psi$ is the anomalous dimension. We determine $\eta_\psi$ by evaluating the leading-order self-energy correction $\Sigma_\psi(k)$ arising from the gauge field (\figref{fig:fermion_self_energy_correction}) and extracting the logarithmically divergent $\slashed{k}\log k$ contribution:
\begin{figure}
    \centering
    \includegraphics[width=0.6\linewidth]{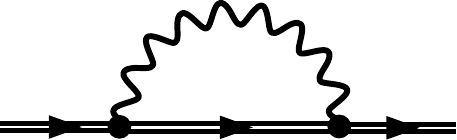}
    \caption{Self-energy correction to the fermion propagator due to the emergent gauge field.}
    \label{fig:fermion_self_energy_correction}
\end{figure}

\begin{equation}
\begin{split}\label{eq:fermion_selfenergy}
    \Sigma_\psi(k)&=\int \frac{\dd^3q}{(2\pi)^3}\gamma_\mu G_\psi(k+q) \gamma_\nu D_{\mu\nu}(-q)\\
    &\quad\rightarrow\frac{8}{(N_f+4N_b)\pi^2}\left(\frac{1}{3}-\zeta\right)\slashed{k}\log k\,.
\end{split}
\end{equation}
Here and in the following, loop integrals are evaluated using dimensional regularization, consistent with the approach in Refs.~\cite{Kaul08,Christos_2024,Feuerpfeil_2026}. This yields the gauge-dependent anomalous dimension
\begin{equation}
    \eta_\psi=\frac{8(1-3\zeta)}{3(N_f+4N_b)\pi^2}\,.
\end{equation}
The gauge dependence above is a reflection of the fact that the fermion Green's function is not gauge-invariant.

\subsection{Boson self-energy}
\begin{figure}
    \centering
    \includegraphics[width=1.0\linewidth]{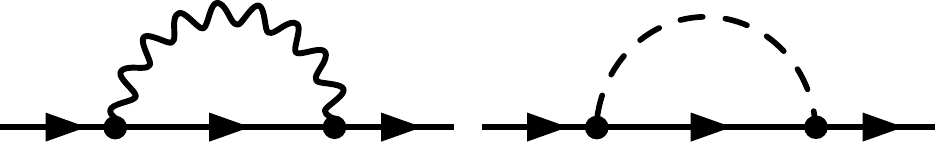}
    \caption{Self-energy corrections to the Higgs propagator arising from fluctuations of the gauge field and the Lagrange multiplier field.}
    \label{fig:boson_self_energy_diagrams}
\end{figure}

The scaling dimension of the Higgs fields $\Phi$ is
\begin{equation}
    \dim[\Phi] = \frac{D-2+\eta_\Phi}{2}\,
\end{equation}
with anomalous dimension $\eta_\Phi$. We compute this at order $\mathcal{O}(1/N_{f,b})$ by extracting the $k^2\log k$ divergence from the diagrams shown in \figref{fig:boson_self_energy_diagrams}.  We begin with the corrections due to the gauge field:
\begin{equation}
\begin{split}\label{eq:I_A1}
    I_{a;1}&=4\int \frac{\dd^3 q}{(2\pi)^3}G_\Phi(k+q)D_{\mu\nu}(-q)(2k+q)_\mu (2k+q)_\nu \\
    &\quad\rightarrow -\frac{16}{(N_f+4N_b)\pi^2}\left(\frac{10}{3}+2\zeta\right)k^2 \log k\,.
\end{split}
\end{equation}
For the Lagrange multiplier fluctuations, we obtain
\begin{equation}
\begin{split}\label{eq:I_lambda1}
    I_{\lambda;1} &= i^2 \int \frac{\dd^3 q}{(2\pi)^3}G_\Phi(k+q)D_\lambda(-q) \rightarrow \frac{4}{3N_b\pi^2}k^2\log k\,.
\end{split}
\end{equation}
Summing these contributions, we conclude that the Higgs anomalous dimension is
\begin{equation}
    \eta_{\Phi}=-\frac{16}{(N_f+4N_b)\pi^2}\left(\frac{10}{3}+2\zeta\right)+\frac{4}{3N_b\pi^2}\,.
\end{equation}

\subsection{Fermion bilinears}
\begin{figure}
    \centering
    \includegraphics[width=0.9\linewidth]{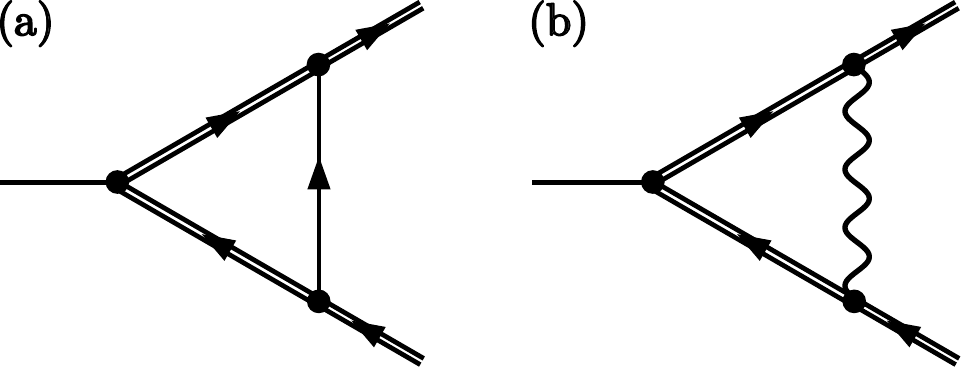}
    \caption{Vertex corrections to the scaling dimensions of the composite fermion operator $\bar{\psi}\sigma^a\mu^b\psi$.}
    \label{fig:composite_vertex_corrections}
\end{figure}

The scaling dimensions of mass terms of the form $\bar{\psi}\sigma^a\mu^b\psi$, where $\sigma$ and $\mu$ act on the spin and valley indices, are given by
\begin{equation}
\dim[\bar{\psi}\sigma^a\mu^b\psi]=2\dim[\psi]+\eta_{\mathrm{vrtx}}\,,
\end{equation}
where $\eta_{\mathrm{vrtx}}$ captures the vertex corrections shown in Figs.~\ref{fig:composite_vertex_corrections}(a) and (b). We can safely neglect the correction arising from the Yukawa coupling [diagram (a)], as it does not contribute a logarithmic divergence at this order.

Therefore, we are left with diagram (b) due to the gauge fluctuations. Taking the limit of vanishing external momenta ($k_1=k_2=0$) allows us to directly extract the logarithmic divergence:
\begin{equation}
\begin{split}
    I_{a;2}&=\int \frac{\dd^3q}{(2\pi)^3}\gamma_\mu G_\psi(q)\sigma^a\mu^b G_\psi(q)\gamma_\nu D_{\mu\nu}(-q)\\
    &\rightarrow -\frac{8\sigma^a\mu^b}{(N_f+4N_b)\pi^2}(3-\zeta)\,.
\end{split}
\end{equation}
Because the gauge interaction involves $\gamma$ matrices that act purely in Dirac space, they commute with $\sigma$ and $\mu$. Consequently, the vertex corrections are identical for all bilinears $\sigma^a\mu^b$, leading to the universal scaling dimension:
\begin{equation}
    \dim[\bar{\psi}\sigma^a\mu^b\psi]=2-\frac{64}{3(N_f+4N_b)\pi^2}\,.
\end{equation}
Note this is independent of $\zeta$, as we are now considering the renormalization of gauge-invariant operators.
Setting $N_b=0$ and $N=N_f/2$, we successfully recover Eq.~(16) of Ref.~\cite{Rantner2002}. As argued in Ref.~\cite{Hermele2008}, for small $N_f$, the gauge fluctuations strongly bind $\bar{\psi}$ and $\psi$, thereby reducing the scaling dimension of the fermion bilinears. We, therefore, expect this dimension to remain below $2$, even away from the strictly large-$N_{f,b}$ limit. 

Crucially, the bosonic contributions increase the scaling dimension of the $\mathrm{SU}(4)$ mass terms and thus reduce their relevance compared to the pure Dirac spin liquid. Because these mass terms correspond to various magnetic and nonmagnetic order parameters~\cite{Hermele2008}, our results indicate that the spatial susceptibilities related to these orders decay as
\begin{equation}
    \chi(r)\propto |r|^{-4+128/(3(N_f+4N_b)\pi^2)}
\end{equation}
within the large-$N_{f,b}$ expansion.

\subsection{Correlation-length exponent}
Following the framework of Ref.~\cite{Kaul08}, we compute the correlation-length exponent $\nu$, defined via the divergence of the correlation length $\xi \propto (g-g_c)^{-\nu}$. This exponent is determined by the scaling relation
\begin{equation}
    \nu = \frac{\gamma_\Phi}{2-\eta_\Phi}\,,
\end{equation}
where $\eta_\Phi$ is the Higgs anomalous dimension and $\gamma_\Phi$ characterizes the scaling of the mass parameter as the critical point is approached:
\begin{equation}
    G_\Phi^{-1}(k=0) \sim (g-g_c)^{\gamma_\Phi}\,.
\end{equation}
To evaluate these quantities, we define the deviation from criticality as $(1/g_c - 1/g) = \sqrt{r_g}/4\pi$. As established in Ref.~\cite{Kaul08}, $\gamma_\Phi$ is related to the coefficient $\alpha$ of the logarithmic divergence $r_g \log(r_g)$ in the regularized self-energy difference,
\begin{equation}
    T(r)\coloneqq \Sigma(0,r) - \frac{\Pi_\lambda(0,0)}{\Pi_\lambda(0,r)}\Sigma(0,0)\,,
\end{equation} 
via
\begin{equation}
    \gamma_\Phi = 2(1-\alpha)\,.
\end{equation}

\begin{figure}[t]
    \centering
    \includegraphics[width=1.0\linewidth]{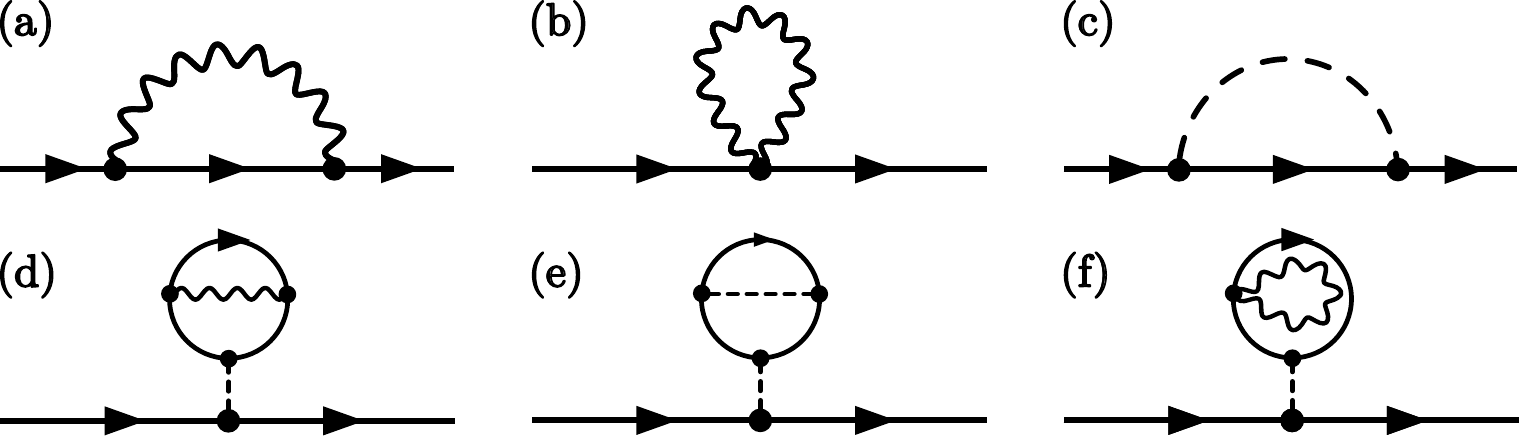}
    \caption{Self-energy diagrams for the Higgs field $\Phi$ at order $1/N_{f,b}$. (a) and (b) are the diagrams due to gauge fluctuations, (c)--(f) due to the Lagrange multiplier fluctuations. At the critical point ($r=0$), only diagrams (a) and (c) contribute to the anomalous dimension $\eta_\Phi$. The full set of diagrams is required to determine $\gamma_\Phi$ and, through scaling relations, the correlation-length exponent $\nu$.}
    \label{fig:correlation_length_contributions}
\end{figure}

The contributing $1/N$ self-energy diagrams are shown in \figref{fig:correlation_length_contributions}. Their analytic expressions are given by
\begin{equation}
\begin{split}
    \Sigma^{(a)} &= I_{a;1}\,,\\ 
    \Sigma^{(b)} &= 4 \sum_{\mu,\nu}\int \frac{\dd^3 q}{(2\pi)^3}\frac{1}{\Pi_a(q)}\left(\delta_{\mu\nu}-\zeta \frac{q_\mu q_\nu}{q^2}\right)\,,\\
    \Sigma^{(c)} &= I_{\lambda;1}\,,\\
    \Sigma^{(d)} &= \frac{i^2}{\Pi_\lambda(0,r)}\int \frac{\dd^3 q}{(2\pi)^3}N_bG_\Phi(q)^2I_{a;1}(q)\,,\\
    \Sigma^{(e)} &= \frac{i^2}{\Pi_\lambda(0,r)}\int \frac{\dd^3 q}{(2\pi)^3}N_bG_\Phi(q)^2I_{\lambda;1}(q)\,,\\
    \Sigma^{(f)} &= \frac{i^2\Sigma^{(b)}}{\Pi_{\lambda}(0,r)}\int \frac{\dd^3 q}{(2\pi)^3}N_bG_\Phi(q)^2 = - \Sigma^{(b)}\,.\\
\end{split}
\end{equation}
Combining the contributions from the Lagrange multiplier fluctuations ($\Sigma = \Sigma^{(c)}+\Sigma^{(e)}$), we find:
\begin{equation}
    \alpha_{\lambda} = \frac{6}{N_b\pi^2}\,.
\end{equation}

For the gauge-mediated contributions ($\Sigma = \Sigma^{(a)}+\Sigma^{(d)}$), the derivation follows the procedure in Ref.~\cite{Kaul08}. However, the charge-2 nature of the Higgs field introduces an overall factor of 4 in the gauge-boson vertices, yielding:
\begin{equation}
\begin{split}
    T(r)&=4\int \frac{q^2 \dd q}{2\pi^2}\Bigg[ \frac{1-\zeta}{\Pi_a(q,r_g)}\frac{q^2}{q^2+r_g}\\ 
    &\quad-\frac{N_b}{\Pi_{\lambda}(0,r_g)}\frac{I_A(q,r_g)}{\Pi_a(q,r_g)}\\
    &\quad+\frac{N_b}{\Pi_\lambda(0,r_g)}\frac{1}{4q\Pi_a(q,0)}\Bigg]\,,
\end{split}
\end{equation}
where $I_A(q,r_g) \approx (1-\zeta) \frac{\Pi_\lambda(0,r_g)}{N_b} + \frac{1}{4q} - \frac{\sqrt{r_g}}{\pi q^2}$. Expanding the integrand for large $q$ isolates the logarithmic divergence:
\begin{equation}
\begin{split}
    \alpha_A=-\frac{16}{(N_f+4N_b)\pi^2}\left(\zeta+\frac{7N_f-4N_b}{N_f+4N_b}\right)\,.
\end{split}
\end{equation}

Summing these terms ($\alpha = \alpha_\lambda + \alpha_A$), we obtain the mass scaling exponent:
\begin{equation}
    \gamma_\Phi=2-\frac{12}{N_b\pi^2}+\frac{32}{\pi^2}\frac{7N_f-4N_b+\zeta(N_f+4N_b)}{(N_f+4N_b)^2}\,.
\end{equation}
Finally, the correlation-length exponent is obtained via $\nu \approx \frac{\gamma_\Phi}{2}(1+\frac{\eta_\Phi}{2})$:
\begin{equation}
\begin{split}
    \nu&\approx 1 - \frac{16}{3N_b\pi^2} + \frac{16}{\pi^2}\frac{7N_f-4N_b}{(N_f+4N_b)^2}-\frac{80}{3(N_f+4N_b)\pi^2}\,.
    \label{eq:nures}
\end{split}
\end{equation}
As expected, the gauge-parameter $\zeta$ cancels exactly, rendering the physical exponent $\nu$ gauge-invariant.

\begin{table*}
    \centering
    \begin{tblr}{
          width = 0.9\linewidth,
          colspec = {c X[c] c c },
          rowsep = 3pt,
          cells = {valign=m},
          cells = {font=\linespread{1.3}\selectfont}
        }\hline\hline
        & Formula & Triangular/Kagome & Maple-leaf \\\hline 
        $(N_f,N_b)$ & $ $ & $ (4,1)$ & $ (12,1)$ \\\hline
        $\dim[\bar{\psi}\sigma^a\mu^b\psi]$ & $ 2-\dfrac{64}{3(N_f+4N_b)\pi^2}$ & $ 1.730$ & $1.865 $ \\
        $\nu$ & $1 - \dfrac{16}{3N_b\pi^2} + \dfrac{16}{\pi^2}\dfrac{7N_f-4N_b}{(N_f+4N_b)^2}-\dfrac{80}{3(N_f+4N_b)\pi^2} $ & $0.730 $ & $ 0.797$ \\
        $\dim[y]$ & $\dfrac{1}{2} - \dfrac{2}{3N_b\pi^2}+\dfrac{160}{3(N_f+4N_b)\pi^2} $ & $1.108 $ & $0.770$\\
        $\dim[y_2]$ & $\dfrac{1}{2} - \dfrac{2}{3N_b\pi^2}+\dfrac{32}{(N_f+4N_b)\pi^2} $ &  & $ 0.635$\\\hline
    \end{tblr}
    \caption{Summary of the scaling dimensions and critical exponents for the corresponding QED$_3$ theories of the gapless $\U(1)$ spin liquids on the triangular, kagome, and maple-leaf lattices in the large-$N_{f,b}$ expansion. Here, $y$ denotes the Dirac-matrix Yukawa coupling: it gaps the pinned triangular/kagome Dirac cones and shifts the unpinned maple-leaf cones. The coupling $y_2$ is the additional maple-leaf coupling without a Dirac gamma matrix.}
    \label{tab:summary_scaling}
\end{table*}

\subsection{Renormalization of the Yukawa coupling}
\label{sec:yukawa}
\begin{figure}[t]
    \centering
    \includegraphics[width=0.9\linewidth]{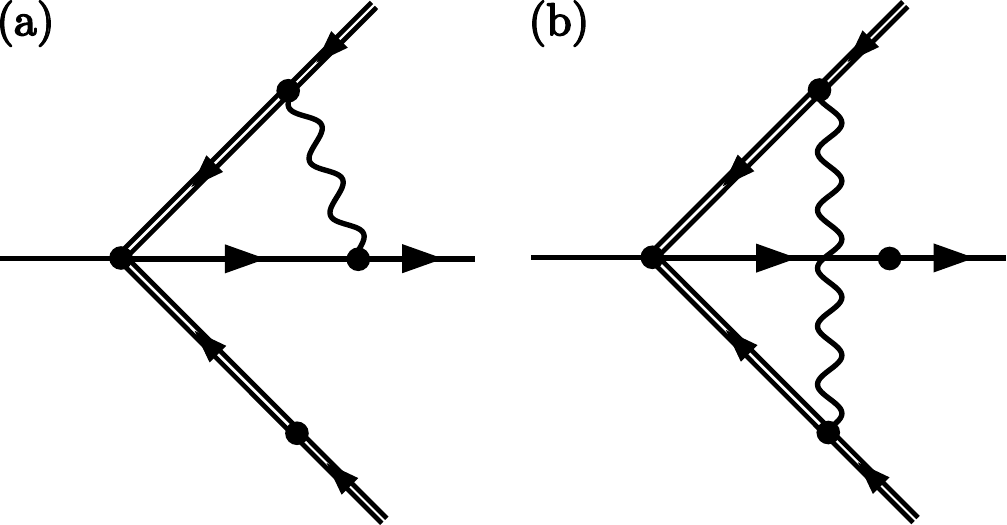}
    \caption{Vertex corrections to the Yukawa coupling. Diagram (a) carries a combinatorial factor of two due to the coupling to either fermion line, whilst diagram (b) only contributes once. We show only the corrections for $\psi^\top \psi$; the Hermitian conjugate receives identical corrections.}
    \label{fig:yukawa_vertex_corrections}
\end{figure}
Finally, we investigate the stability of the Higgs critical point by calculating the vertex corrections to the scaling dimension of the Yukawa coupling $y$ and $y_2$. The scaling dimension of a generic Yukawa operator is given by
\begin{equation}\label{eq:dim_yukawa}
    \dim[\Phi^* \psi^\top X\psi]=2\dim[\psi]+\dim[\Phi]+\eta_{\text{vrtx}}\,,
\end{equation}
where $\eta_{\text{vrtx}}$ captures the vertex corrections due to gauge and Lagrange multiplier fluctuations. The scaling dimension of the coupling constant itself is then $\dim[y] = 3 - \dim[\text{operator}]$.

The contributing diagrams are shown in \figref{fig:yukawa_vertex_corrections}. The first diagram [\figref{fig:yukawa_vertex_corrections}(a)] carries a combinatorial factor of two, as the boson can couple to either of the two fermion lines. Evaluating these diagrams at vanishing external momenta to isolate the logarithmic divergences, we find for the first diagram:
\begin{equation}\label{eq:yukawa_renormalization_1}
\begin{split}
    I_{\Phi\psi} &= -2i^2X\int \frac{\dd^3 q}{(2\pi)^3}\gamma_\mu G_\psi(q) G_\Phi(q)q_\nu D_{\mu\nu}(-q) \\
    &=\frac{32 X}{N_f+4N_b}\int \frac{\dd^3 q}{(2\pi)^3}\frac{(1-\zeta)}{q^3}\\
    &\rightarrow-\frac{16 X}{(N_f+4N_b)\pi^2}(1-\zeta)\log k\,.
\end{split}
\end{equation}
For the second diagram [\figref{fig:yukawa_vertex_corrections}(b)], we distinguish two cases. First, if $X$ does not involve a Dirac gamma matrix---as for the secondary maple-leaf coupling $y_2$ in \eqnref{eq:yukawa_ml}---one obtains:
\begin{equation}\label{eq:yukawa_renormalization_2}
\begin{split}
    I_{\psi\psi}^{(a)} &= X\int \frac{\dd^3 q}{(2\pi)^3}\gamma_\mu G_\psi(q) G_\psi(-q)\gamma_\nu D_{\mu\nu}(-q)\\
    &=-\frac{16X}{N_f+4N_b} \int \frac{\dd^3 q}{(2\pi)^3}\gamma_\mu \frac{1}{q^3}\gamma_\nu\left(\delta_{\mu\nu}-\zeta\frac{q_\mu q_\nu}{q^2}\right)\\
    &=-\frac{16X}{N_f+4N_b} \int \frac{\dd^3 q}{(2\pi)^3}\frac{(3-\zeta)}{q^3}\\
    &\rightarrow \frac{8X}{(N_f+4N_b)\pi^2}(3-\zeta)\log k\,.
\end{split}
\end{equation}
Second, if $X\propto \sum_{a=0,x,y}c_a\gamma^a$ contains a Dirac-index matrix---as for the triangular/kagome Yukawa coupling in \eqnref{eq:ml_yukawa} and for the primary maple-leaf coupling $y$ in \eqnref{eq:yukawa_ml}---the correction is given by
\begin{equation}\label{eq:yukawa_2_renormalization_2}
\begin{split}
    I_{\psi\psi}^{(b)} &= \int \frac{\dd^3 q}{(2\pi)^3}\gamma_\mu G_\psi(q) X G_\psi(-q)\gamma_\nu D_{\mu\nu}(-q)\\
    &\rightarrow \frac{8X}{(N_f+4N_b)\pi^2}\left(\frac{1}{3}-\zeta\right)\log k\,.
\end{split}
\end{equation}
Combining these with the previously computed anomalous dimensions for the fields, and using the Dirac-matrix result for $y$ and the Dirac-trivial result for $y_2$, we obtain the scaling dimensions of the coupling constants:

\begin{equation}
\begin{split}
    \dim[y] &= \frac{1}{2} - \frac{2}{3N_b\pi^2}+\frac{160}{3(N_f+4N_b)\pi^2}\,,\label{eq:yres}\\
    \dim[y_2]&=\frac{1}{2}-\frac{2}{3N_b\pi^2}+\frac{32}{(N_f+4N_b)\pi^2} \,.
\end{split}
\end{equation}
The result in \eqnref{eq:yres} reveals a competition between the different matter sectors: while fluctuations of the Lagrange multiplier $\lambda$ tend to reduce the scaling dimension of the coupling constant (the second term), the gauge field fluctuations act to increase it (the third term).

The instability of the Higgs fixed point toward symmetry-allowed Yukawa perturbations
can be phrased explicitly as a renormalization-group statement. At leading order
in $1/N$, the scaling dimension $\dim[y]$ (and similarly $\dim[y_2]$) extracted in \eqnref{eq:yres} governs the flow of the
Yukawa coupling $y$ via
\begin{equation}
    \frac{\dd y}{\dd \ell}=\dim[y] y+\mathcal{O}(y^3)\,,
\end{equation}
where $\ell = \ln(\Lambda_0/\Lambda)$ is the RG ``time'' associated with the rescaling of the momentum cutoff $\Lambda$~\cite{Wilson_1973}. Since $\dim[y] > 0$ (and likewise $\dim[y_2] > 0$ where the latter is present) for the physical case $N_b = 1$, both couplings are relevant perturbations of the Higgs fixed point and flow to strong coupling in the infrared.

We emphasize that the large-$N_{f,b}$ expansion does not, by itself, control the RG flow at the physical values $(N_f, N_b)$ of these strongly coupled gauge theories; rather, it reliably identifies the symmetry-allowed operator content~\cite{Sachdev2018} and the sign of the leading contribution to $\dim[y]$. For the maple-leaf lattice, $N_f = 12$ is large enough that
subleading corrections are parametrically small, $\mathcal{O}(1/N_f^2) \sim 1\%$, but
since $N_b = 1$ is not itself large, the extrapolation to the physical boson sector is not
quantitatively controlled. The relevance of $y$ and $y_2$ established above should, therefore, be understood as a leading-order large-$N$ result identifying the correct sign
and operator content, rather than a fully resolved nonperturbative RG flow.

\subsection{Flavor-dependent scaling and fixed-point stability}

We conclude by summarizing the computed scaling dimensions and critical exponents for the various lattices. The continuum theories for the triangular and kagome lattices are identical, while the maple-leaf lattice is distinguished by its larger fermion flavor number ($N_f=12$) and the presence of an additional Yukawa coupling. The numerical values for the physically relevant case of $N_b=1$ are presented in \tabref{tab:summary_scaling}.

As shown in \tabref{tab:summary_scaling}, the physical case of $N_b=1$ yields a Yukawa scaling dimension of $\dim[y] \approx 1.108$ for the $N_f=4$ theory governing the triangular and kagome lattices. On the maple-leaf lattice, the larger fermion flavor number ($N_f=12$) suppresses these fluctuations and substantially reduces the relevance of the primary coupling to $\dim[y]\approx 0.770$. 

Microscopically, the Dirac-matrix Yukawa coupling $y$ has two different lattice realizations. On the triangular and kagome lattices it is the Higgs perturbation that gaps the pinned Dirac cones. On the maple-leaf lattice the same type of Dirac-index coupling is tied to the unpinned nature of the Dirac nodes: the uniform $\mathrm{U}(1)$ state possesses a parametric freedom that allows the positions of the Dirac points to shift continuously along the $\overline{\Gamma \mathrm{M}}$ lines. Thus the triangular/kagome and maple-leaf $y$ couplings have the same leading scaling eigenvalue for a fixed $N_f$, but the maple-leaf perturbation is less relevant numerically because the maple-leaf continuum theory has the larger flavor number $N_f=12$. 

\section{Outlook}

The comparative analysis presented here offers a broader perspective on Higgs transitions out of Dirac spin liquids. For the depleted triangular lattices studied in this work, the overarching structure of the $\text{QED}_3$-Higgs theory remains robust, while the primary quantitative distinction arises from the low-energy flavor content. Specifically, the triangular and kagome lattices realize the same $N_f=4$ theory, whereas the maple-leaf lattice yields an analogous theory with $N_f=12$. Because the resulting Yukawa instability is parametrically weaker in the maple-leaf case, this geometry may provide a favorable setting for extended intermediate-scale Higgs critical behavior. However, the maple-leaf lattice also illustrates a critical trade-off: its Dirac cones are not pinned at high-symmetry momenta but instead move along symmetry-related lines, and this geometric freedom introduces a symmetry-allowed Yukawa coupling, which is more relevant than $y_2$. Our results, therefore, suggest a clear objective for future model-building---namely, the identification of frustrated lattices that combine a large fermion flavor number with symmetry-pinned Dirac nodes, which are not gapped out by pairing terms. Such systems would maximize the stabilizing effect of large $N_f$ while avoiding the destabilizing influence of the additional relevant perturbations associated with nodal mobility.

More broadly, these results show how microscopically distinct lattices, with different site structures and PSG embeddings, can nevertheless flow to a common continuum description. In this sense, our findings establish an emergent unification, analogous in spirit to the unified field-theoretic description of the $\SU(2)$ $\pi$-flux state across the square, Shastry-Sutherland, and checkerboard lattices~\cite{Feuerpfeil_2026,Maity-2026}.

This perspective suggests several natural directions for future work. On the microscopic side, it will be important to test the stability of these candidate $\U(1)$ and proximate $\mathbb{Z}_2$ spin liquids within concrete lattice Hamiltonian models, identifying precisely when distinct microscopic realizations support identical continuum Higgs theories. On the field-theory front, it will also be valuable to extend our calculations beyond the leading large-$N_{f,b}$ approximation to sharpen the quantitative fate of the Yukawa perturbation and to reliably determine the extent of pseudocritical scaling in physically relevant regimes. More generally, the problem addressed here fits into a wider effort to understand weakly unstable fermionic fixed points: the central question is not merely whether they survive asymptotically, but how their long crossover scales shape intermediate-energy physics, competing orders, and observable phenomena in realistic quantum materials.

A further direction involves applying this analytic framework to other frustrated geometries hosting Dirac spin liquids and their paired Higgs descendants. From this broader vantage point, the focus shifts from individual lattice models to identifying which aspects of the Higgs critical theory are universally fixed by the continuum gauge structure, and which are hypersensitive to the low-energy embedding of lattice symmetries, flavor multiplicities, and pairing channels. Finally, it would be interesting to revisit these questions in the broader context of moir\'e triangular systems. Recent experiments on twisted bilayer WSe$_2$ have revealed correlated insulating regimes proximate to superconductivity, suggesting a tunable setting in which insulating and paired phases can occur near one another. We do not make a direct material-specific claim for these systems. Rather, they motivate the broader theoretical problem of understanding how spin-liquid fractionalization, spinon pairing, and Higgs criticality may arise in correlated triangular-lattice settings.

\section{Acknowledgments}
This work is supported by the Deutsche Forschungsgemeinschaft (DFG, German Research Foundation) through Project-ID 258499086---SFB 1170 and through the W\"urzburg-Dresden Cluster of Excellence on Complexity and Topology in Quantum Matter--ctd.qmat Project-ID 390858490---EXC 2147. The Flatiron Institute is a division of the Simons Foundation. S.S. was supported by the U.S. National Science Foundation Grant No. DMR 2245246 and by the Simons Collaboration on Ultra-Quantum Matter which is a grant from the Simons Foundation (651440, S.S.). The work of Y.I. was performed in part at the Aspen Center for Physics, which is supported by a grant from the Simons Foundation (1161654, Troyer). This research was also supported in part by grant NSF PHY-2309135 to the Kavli Institute for Theoretical Physics and by the International Centre for Theoretical Sciences (ICTS) for participating in the Discussion Meeting---Fractionalized Quantum Matter (code: ICTS/DMFQM2025/07). R.T. thanks IIT Madras for a Visiting Faculty Fellow position under the IoE program.  Y.I. acknowledges support from the Abdus Salam International Centre for Theoretical Physics through the Associates Programme, from the Simons Foundation through Grant No.~284558FY19, from IIT Madras through the Institute of Eminence program for establishing QuCenDiEM (Project No. SP22231244CPETWOQCDHOC).

\appendix

\section{Details of the \texorpdfstring{$\pi$}{pi}-flux U(1) and descendant \texorpdfstring{$\mathbb{Z}_2$}{Z2} spin liquid \texorpdfstring{Ansätze}{{\it Ansätze}} on the triangular lattice}\label{app:ansatz_triangular}
\begin{figure*}
    \centering    \includegraphics[width=1.0\linewidth]{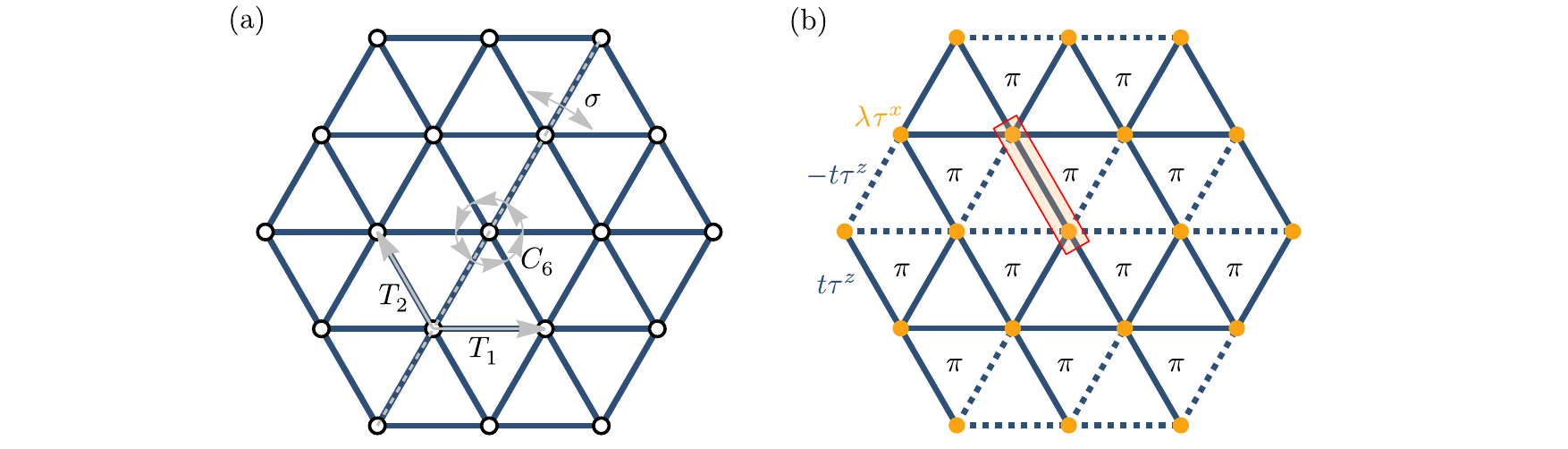}
    \caption{(a) Triangular lattice with its minimal set of space group symmetry generators. (b) Graphical illustration of the $\mathbb{Z}_2$ QSL {\it Ansatz} derived from the $\pi$-flux $\U(1)$ parent state. The parent $\U(1)$ state is recovered by setting the onsite pairing parameter to zero ($\lambda=0$). Realizing this {\it Ansatz} requires doubling the unit cell along the $T_2$ direction, resulting in the two-site unit cell indicated by the red rectangle.}
    \label{fig:z2_qsl_unification}
\end{figure*}
We define the site positions on the triangular lattice using the basis:
\begin{equation}
    \mathbf{r}\equiv(x,y)=x T_1+yT_2\,.
\end{equation}
The triangular lattice is characterized by the wallpaper group $p6mm$, whose elements are generated by two translations ($T_1$ and $T_2$), a six-fold rotation axis ($C_6$) centered at the origin, and a reflection ($\sigma$) across the line $y=\sqrt{3}x$. Additionally, we consider time-reversal symmetry $\mathcal{T}$, which leaves the spatial coordinates $(x,y)$ invariant, see \figref{fig:z2_qsl_unification}. The actions of the spatial generators on the coordinates are defined as follows:
\begin{equation}
\left.\begin{aligned}
T_1:(x,y)&\rightarrow(x+1,y)\,,\\
T_2:(x,y)&\rightarrow(x,y+1)\,,\\
C_6:(x,y)&\rightarrow(x-y,x)\,,\\
\sigma:(x,y)&\rightarrow(y,x)\,.
\end{aligned}\right.
\end{equation}
The symmetry properties of the spinon mean-field states are described by the projective symmetry group (PSG), represented by the set $\{W^{}_{T_1},W^{}_{T_2},W^{}_{C_6},W^{}_{\sigma},W^{}_{\mathcal{T}}\}$. We first define the $\U(1)$ $\pi$-flux {\it Ansatz} via the following link fields:
\begin{equation}
\begin{split}
&u^{}_{(x,y),(x,y+1)}=t\tau^z\,,\\
&u^{}_{(x,y),(x+1,y)}=(-1)^{y+1}t\tau^z\,,\\
&u^{}_{(x,y),(x+1,y+1)}=(-1)^{y+1}t\tau^z\,.\label{eq:trian_u1}
\end{split}
\end{equation}
The corresponding PSG representations for this $\U(1)$ state are given by
\begin{equation}
\begin{split}
W^{}_{T_1}(x,y)&=e^{{i}\phi^{}_{T_1}\tau^z}\,,\\
W^{}_{T_2}(x,y)&=(-1)^{x}e^{{i}\phi^{}_{T_2}\tau^z}\,,\\
W^{}_{C_6}(x,y)&=(-1)^{xy+\frac{y(y-1)}{2}}e^{{i}\phi^{}_{C_6}\tau^z}{i}\tau^x\,,\\
W^{}_{\sigma}(x,y)&=(-1)^{xy}e^{{i}\phi^{}_{\sigma}\tau^z}{i}\tau^x\,,\\
W^{}_{\mathcal{T}}(x,y)&=e^{{i}\phi^{}_{\mathcal{T}}\tau^z}{i}\tau^x\,.\label{eq:trian_u1_Psg}
\end{split}
\end{equation}
A symmetric $\mathbb{Z}_2$ descendant state is obtained by introducing an onsite pairing term:
\begin{equation}
u^{}_{(x,y),(x,y)}=\lambda\tau^x\,.\label{eq:trian_z2_pairing}
\end{equation}
Enforcing consistency with the lattice and time-reversal symmetries requires fixing the $\mathrm{U}(1)$ PSG phases to $\phi^{}_{T_1}=\phi^{}_{T_2}=\phi^{}_{C_6}=\phi^{}_{\sigma}=0$ and $\phi^{}_{\mathcal{T}}=\pi/2$. Under these constraints, the associated $\mathbb{Z}_2$ PSG is defined as
\begin{equation}
\begin{split}
W^{}_{T_1}(x,y)&=\tau^0\,,\\
W^{}_{T_2}(x,y)&=(-1)^{x}\tau^0\,,\\
W^{}_{C_6}(x,y)&=(-1)^{xy+\frac{y(y-1)}{2}}{i}\tau^x\,,\\
W^{}_{\sigma}(x,y)&=(-1)^{xy}{i}\tau^x\,,\\
W^{}_{\mathcal{T}}(x,y)&={i}\tau^y\,.\label{eq:trian_z2_psg}
\end{split}
\end{equation}
\section{Details of the \texorpdfstring{$[0,\pi]$}{[0,pi]}-flux U(1) and descendant \texorpdfstring{$\mathbb{Z}_2$}{Z2} spin liquid \texorpdfstring{Ansätze}{{\it Ansätze}} on the kagome lattice}\label{app:kagome}
\begin{figure*}
    \centering    \includegraphics[width=1.0\linewidth]{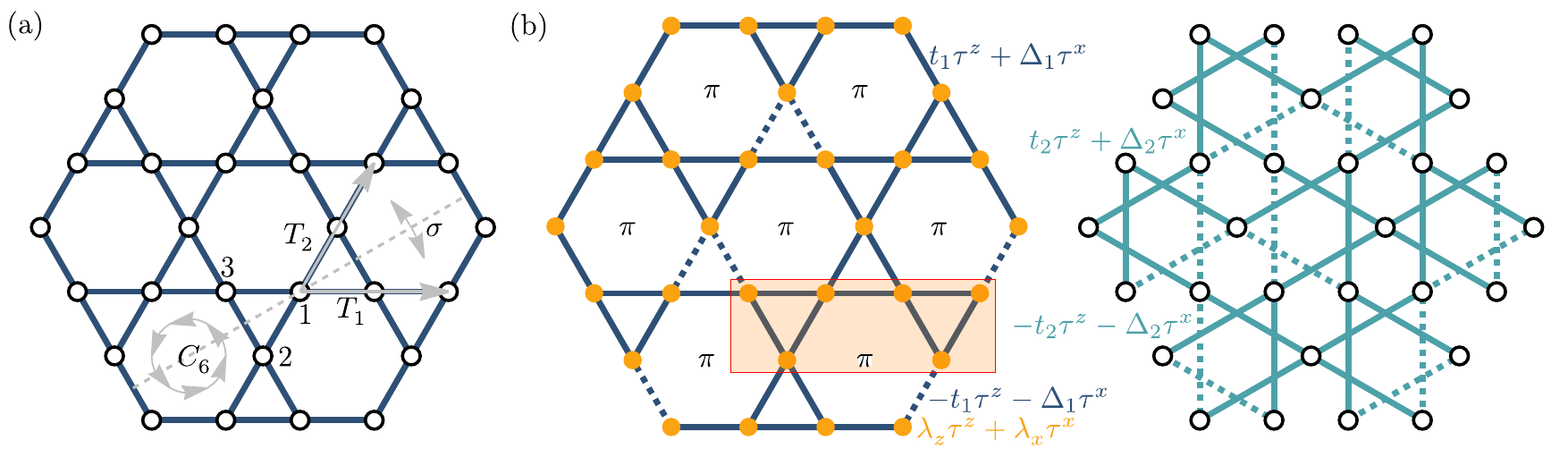}
    \caption{(a) Kagome lattice with its minimal set of space group symmetry generators. (b) Graphical representation of the $\mathbb{Z}_2[0,\pi]\beta$ QSL {\it Ansatz} derived from its $[0,\pi]$-flux $\U(1)$ parent state. The parent $\U(1)$ state is recovered by setting all pairing parameters to zero ($\lambda_x = \Delta_1 = \Delta_2 = 0$). To implement this {\it Ansatz} in a translationally invariant gauge, the unit cell is doubled along the $T_1$ direction, resulting in the six-site unit cell indicated by the red rectangle.} \label{fig:z2_qsl_kagome}
\end{figure*}
We denote the position of each site on the kagome lattice as $(\mathbf{r},s)$, where $s \in \{1,2,3\}$ labels the three sublattices within the unit cell:
\begin{equation}
    (\mathbf{r},s)\equiv(x,y,s)=x T_1+yT_2+\mathbf{r}_s\,.
\end{equation}
Like the triangular lattice, the kagome lattice belongs to the $p6mm$ wallpaper group. Its symmetry elements are generated by two translations ($T_1$ and $T_2$), a six-fold rotation axis ($C_6$) at the origin, and a reflection ($\sigma$) across the line $x=\sqrt{3}y$, as illustrated in \figref{fig:z2_qsl_kagome}(a). The actions of these generators on the site coordinates $(x,y,s)$ are given by
\begin{equation}
\left.\begin{aligned}
T_1:(x,y,s)&\rightarrow(x+1,y,s)\,,\\
T_2:(x,y,s)&\rightarrow(x,y+1,s)\,,\\
C_6:(x,y,s)&\rightarrow(-y-\delta^{}_{s,1}-\delta^{}_{s,3},x+y+\delta^{}_{s,1},C_6(s))\,,\\
\sigma:(x,y,s)&\rightarrow(y,x,\sigma(s))\,.
\end{aligned}\right.
\end{equation}
In these expressions, the sublattice indices $s$ undergo permutations defined by the mapping $C_6(s)=(2,3,1)$ and $\sigma(s)=(1,3,2)$.

We first define the $\mathrm{U}(1)$ $[0,\pi]$-flux {\it Ansatz}. The mean-field link fields and the onsite terms are parameterized as follows:
\begin{equation}
\begin{split}
&u^{}_{(x,y,1),(x,y,2)}=u^{}_{(x,y,2),(x,y,3)}=t^{}_{1}\tau^z\,,\\
&u^{}_{(x,y,3),(x,y,1)}=u^{}_{(x,y,3),(x-1,y,1)}=t^{}_{1}\tau^z\,,\\
&u^{}_{(x,y,2),(x,y-1,1)}=u^{}_{(x,y,2),(x+1,y-1,3)}=(-1)^xt^{}_{1}\tau^z\,,\\
&u^{}_{(x-1,y,1),(x,y,2)}=u^{}_{(x-1,y,2),(x,y,3)}=t^{}_{2}\tau^z\,,\\
&u^{}_{(x,y,2),(x,y-1,3)}=u^{}_{(x,y,3),(x,y-1,1)}=(-1)^xt^{}_{2}\tau^z\,,\\
&u^{}_{(x,y-1,3),(x-1,y,1)}=u^{}_{(x,y-1,1),(x-1,y,2)}=(-1)^{x+1}t^{}_{2}\tau^z\,,\\
&u^{}_{(x,y,s),(x,y,s)}=\lambda^{}_{z}\tau^z\,.\label{eq:kagome_u1}
\end{split}
\end{equation}
The corresponding PSGs are given by
\begin{equation}
\begin{split}
W^{}_{T_1}(x,y,s)&=(-1)^{y}e^{{i}\phi^{}_{T_1}\tau^z}\,,\\
W^{}_{T_2}(x,y,s)&=e^{{i}\phi^{}_{T_2}\tau^z}\,,\\
W^{}_{C_6}(x,y,s)&=(-1)^{xy+\frac{x(x+1)}{2}+(x+y)\delta^{}_{s,3}}e^{{i}\phi^{}_{C_6}\tau^z}\,,\\
W^{}_{\sigma}(x,y,s)&=(-1)^{xy}e^{{i}\phi^{}_{\tau}\tau^z}\,,\\
W^{}_{\mathcal{T}}(x,y,s)&=e^{{i}\phi^{}_{\mathcal{T}}\tau^z}{i}\tau^x\,\,.\label{eq:kagome_u1_PSG}
\end{split}
\end{equation}
A descendant $\mathbb{Z}_2$ state, designated as $\mathbb{Z}_2[0,\pi]\beta$, is constructed by introducing bond and onsite pairing terms, which correspond to replacing
\begin{equation}
\begin{split}
    t_1\tau^z&\rightarrow t_1\tau^z+\Delta_1 \tau^x\,,\\
    t_2\tau^z&\rightarrow t_2\tau^z+\Delta_2\tau^x\,,\\
    \lambda_z\tau^z&\rightarrow \lambda_z\tau^z+\lambda_x\tau^x
\end{split}
\end{equation}
in \eqnref{eq:kagome_u1}.
Under this pairing perturbation, the invariant gauge group is reduced from $\U(1)$ to $\mathbb{Z}_2$, and the resulting PSG representations are
\begin{equation}
\begin{split}
W^{}_{T_1}(x,y,s)&=(-1)^{y}\tau^0\,,\\
W^{}_{T_2}(x,y,s)&=\tau^0\,,\\
W^{}_{C_6}(x,y,s)&=(-1)^{xy+\frac{x(x+1)}{2}+(x+y)\delta^{}_{s,3}}\tau^0\,,\\
W^{}_{\sigma}(x,y,s)&=(-1)^{xy}\tau^0\,,\\
W^{}_{\mathcal{T}}(x,y,s)&={i}\tau^y\,\,.\label{eq:kagome_z2_PSG}
\end{split}
\end{equation}
\section{Details of the uniform U(1) and descendant \texorpdfstring{$\mathbb{Z}_2$}{Z2} spin liquid \texorpdfstring{Ansätze}{{\it Ansätze}} on the maple-leaf lattice}
\label{app:mapleleaf}
\begin{figure*}
    \centering    \includegraphics[width=1.0\linewidth]{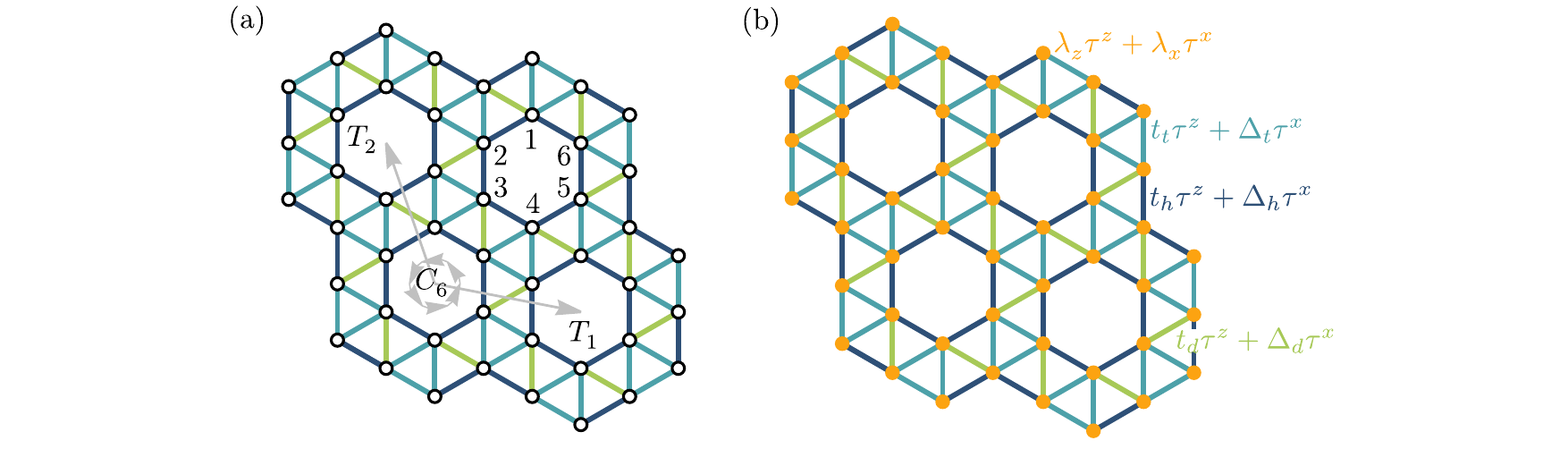}
    \caption{(a) Maple-leaf lattice with its minimal set of space group symmetry generators. (b) Graphical illustration of the $\mathbb{Z}_2$ QSL {\it Ansatz} (labeled Z0002) derived from the uniform $\mathrm{U}(1)$ parent state (labeled UC00). The parent state is recovered by setting all pairing parameters to zero ($\lambda_x = \Delta_h = \Delta_d = \Delta_t = 0$).}    \label{fig:z2_qsl_maple}
\end{figure*}
We define the unit cell as consisting of six sites, indexed by $s=1, \dots, 6$. The position of any site is denoted by $(x,y,s)$, where the coordinates $(x,y)$ specify the unit cell location.
The maple-leaf lattice belongs to the $p6$ wallpaper group, which is generated by two translations ($T_1$ and $T_2$) and a six-fold rotation ($C_6$) centered at the origin, as illustrated in \figref{fig:z2_qsl_maple}(a). The actions of these generators on the site coordinates $(x,y,s)$ are given by
\begin{equation}
\left.\begin{aligned}
T_1:(x,y,s)&\rightarrow(x+1,y,s)\,,\\
T_2:(x,y,s)&\rightarrow(x,y+1,s)\,,\\
C_6:(x,y,s)&\rightarrow(x-y,x,C_6(s))\,.
\end{aligned}\right.
\end{equation}
Under the $C_6$ rotation, the sublattice indices are cyclically permuted according to the mapping $C_6(s) = (2, 3, 4, 5, 6, 1)$. The uniform $\mathrm{U}(1)$ {\it Ansatz}, called UC00 in Ref.~\cite{Sonnenschein-2024}, is defined by the following link and onsite fields:
\begin{equation}
\begin{split}
u^{}_{\text{hex}}&=t^{}_{h}\tau^z\,,\\
u^{}_{\text{dim}}&=t^{}_{d}\tau^z\,,\\
u^{}_{\text{tria}}&=t^{}_{t}\tau^z\,,\\
u^{}_{\text{onsite}}&=\lambda^{}_{z}\tau^z\,.\label{eq:maple_u1}
\end{split}
\end{equation}
The corresponding PSGs are given by
\begin{equation}
\begin{split}
W^{}_{T_1}(x,y,s)&=e^{{i}\phi^{}_{T_1}\tau^z}\,,\\
W^{}_{T_2}(x,y,s)&=e^{{i}\phi^{}_{T_2}\tau^z}\,,\\
W^{}_{C_6}(x,y,s)&=e^{{i}\phi^{}_{C_6}\tau^z}\,,\\
W^{}_{\mathcal{T}}(x,y,s)&=e^{{i}\phi^{}_{\mathcal{T}}\tau^z}{i}\tau^x\,\,.\label{eq:maple_u1_PSG}
\end{split}
\end{equation}
The descendant uniform $\mathbb{Z}_2$ state (Z0002 in Ref.~\cite{Sonnenschein-2024}) is constructed by adding uniform pairing terms to the hexagon, dimer, and triangle bonds, alongside an onsite pairing term. This leads to the mean-field:
\begin{equation}
\begin{split}
u^{}_{\text{hex}}&=t^{}_{h}\tau^z+\Delta^{}_{h}\tau^x\,,\\
u^{}_{\text{dim}}&=t^{}_{d}\tau^z+\Delta^{}_{d}\tau^x\,,\\
u^{}_{\text{tria}}&=t^{}_{t}\tau^z+\Delta^{}_{t}\tau^x\,,\\
u^{}_{\text{onsite}}&=\lambda^{}_{z}\tau^z+\lambda^{}_{x}\tau^x\,.\label{eq:maple_z2}
\end{split}
\end{equation}
The corresponding PSGs are given by
\begin{equation}
\begin{split}
W^{}_{T_1}(x,y,s)&=\tau^0\,,\\
W^{}_{T_2}(x,y,s)&=\tau^0\,,\\
W^{}_{C_6}(x,y,s)&=\tau^0\,,\\
W^{}_{\mathcal{T}}(x,y,s)&={i}\tau^y\,\,.\label{eq:maple_z2_psg}
\end{split}
\end{equation}

\begin{figure*}
    \centering    \includegraphics[width=1.0\linewidth]{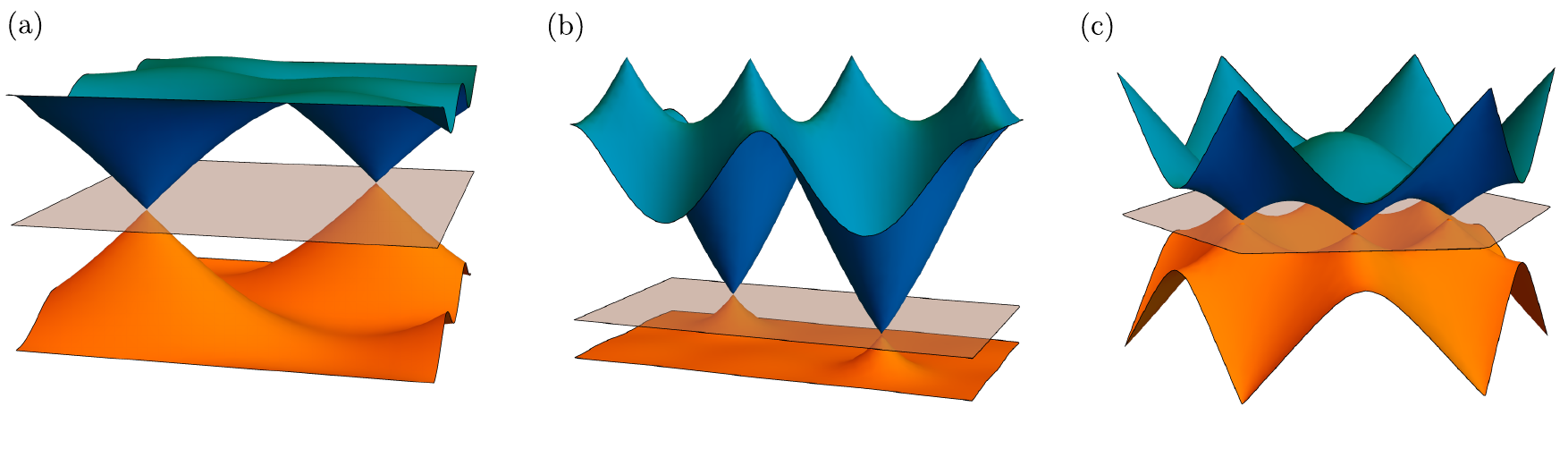}
    \caption{(a) Dirac dispersion of the $\pi$-flux $\mathrm{U}(1)$ state on the triangular lattice; Dirac nodes appear at $\pm(\pi/2, -\pi//2\sqrt{3}))$. (b) Dirac dispersion of the $[0,\pi]$-flux $\mathrm{U}(1)$ state on the kagome lattice ($t_2/t_1=0.1$); Dirac nodes appear at $\pm(\pi/2, \pi/(2\sqrt{3}))$. (c) Dirac dispersion of the uniform $\mathrm{U}(1)$ state on the maple-leaf lattice ($t_h/t_d = -0.75$, $t_t/t_d = 0$). The Dirac nodes are located on the $\overline{\Gamma \mathrm{M}}$ lines, where the $\mathrm{M}$ points are given by the set $\left\{ \pm\left(\frac{\pi}{7\sqrt{3}}, \frac{3\pi}{7}\right), \pm\left(-\frac{4\pi}{7\sqrt{3}}, \frac{2\pi}{7}\right), \pm\left(-\frac{5\pi}{7\sqrt{3}}, -\frac{\pi}{7}\right) \right\}$.}
    \label{fig:U1_dirac_qsl}
\end{figure*}

The spinon dispersions of the three lattices are shown in \figref{fig:U1_dirac_qsl}.

\bibliography{refs.bib}

@article{He-2017,
  title = {{Signatures of Dirac Cones in a DMRG Study of the Kagome Heisenberg Model}},
  author = {He, Yin-Chen and Zaletel, Michael P. and Oshikawa, Masaki and Pollmann, Frank},
  journal = {Phys. Rev. X},
  volume = {7},
  issue = {3},
  pages = {031020},
  numpages = {16},
  year = {2017},
  month = {Jul},
  publisher = {American Physical Society},
  doi = {10.1103/PhysRevX.7.031020},
  url = {https://link.aps.org/doi/10.1103/PhysRevX.7.031020}
}

@article{Wen2002,
  title = {{Quantum orders and symmetric spin liquids}},
  author = {Wen, Xiao-Gang},
  journal = {Phys. Rev. B},
  volume = {65},
  issue = {16},
  pages = {165113},
  numpages = {37},
  year = {2002},
  month = {Apr},
  publisher = {American Physical Society},
  doi = {10.1103/PhysRevB.65.165113},
  url = {https://link.aps.org/doi/10.1103/PhysRevB.65.165113}
}

@article{Ran-2007,
  title = {{Projected-Wave-Function Study of the Spin-$1/2$ Heisenberg Model on the Kagom\'e Lattice}},
  author = {Ran, Ying and Hermele, Michael and Lee, Patrick A. and Wen, Xiao-Gang},
  journal = {Phys. Rev. Lett.},
  volume = {98},
  issue = {11},
  pages = {117205},
  numpages = {4},
  year = {2007},
  month = {Mar},
  publisher = {American Physical Society},
  doi = {10.1103/PhysRevLett.98.117205},
  url = {https://link.aps.org/doi/10.1103/PhysRevLett.98.117205}
}

@article{Senthil_2004,
author = {T. Senthil  and Ashvin Vishwanath  and Leon Balents  and Subir Sachdev  and Matthew P. A. Fisher },
title = {{Deconfined Quantum Critical Points}},
journal = {Science},
volume = {303},
number = {5663},
pages = {1490-1494},
year = {2004},
doi = {10.1126/science.1091806},
URL = {https://www.science.org/doi/abs/10.1126/science.1091806},
eprint = {https://www.science.org/doi/pdf/10.1126/science.1091806}}

@article{Senthil2004a,
  title = {{Quantum criticality beyond the Landau-Ginzburg-Wilson paradigm}},
  author = {Senthil, T. and Balents, Leon and Sachdev, Subir and Vishwanath, Ashvin and Fisher, Matthew P. A.},
  journal = {Phys. Rev. B},
  volume = {70},
  issue = {14},
  pages = {144407},
  numpages = {33},
  year = {2004},
  month = {Oct},
  publisher = {American Physical Society},
  doi = {10.1103/PhysRevB.70.144407},
  url = {https://link.aps.org/doi/10.1103/PhysRevB.70.144407}
}

@article{Sachdev2018,
doi = {10.1088/1361-6633/aae110},
url = {https://doi.org/10.1088/1361-6633/aae110},
year = {2018},
month = {nov},
publisher = {IOP Publishing},
volume = {82},
number = {1},
pages = {014001},
author = {Sachdev, Subir},
title = {{Topological order, emergent gauge fields, and Fermi surface reconstruction}},
journal = {Rep. Prog. Phys.},
}

@article{Wilson_1973,
    author = "Wilson, K. G. and Kogut, John B.",
    title = "{The Renormalization group and the epsilon expansion}",
    doi = "10.1016/0370-1573(74)90023-4",
    journal = "Phys. Rept.",
    volume = "12",
    pages = "75--199",
    year = "1974"
}

@article{Affleck1988,
  title = {{Large-n limit of the Heisenberg-Hubbard model: Implications for high-${T}_{c}$ superconductors}},
  author = {Affleck, Ian and Marston, J. Brad},
  journal = {Phys. Rev. B},
  volume = {37},
  issue = {7},
  pages = {3774--3777},
  numpages = {0},
  year = {1988},
  month = {Mar},
  publisher = {American Physical Society},
  doi = {10.1103/PhysRevB.37.3774},
  url = {https://link.aps.org/doi/10.1103/PhysRevB.37.3774}
}

@article{Dagotto-1988,
  title = {{SU(2) gauge invariance and order parameters in strongly coupled electronic systems}},
  author = {Dagotto, Elbio and Fradkin, Eduardo and Moreo, Adriana},
  journal = {Phys. Rev. B},
  volume = {38},
  issue = {4},
  pages = {2926--2929},
  numpages = {0},
  year = {1988},
  month = {Aug},
  publisher = {American Physical Society},
  doi = {10.1103/PhysRevB.38.2926},
  url = {https://link.aps.org/doi/10.1103/PhysRevB.38.2926}
}

@article{Abrikosov-1965,
  title = {{Electron scattering on magnetic impurities in metals and anomalous resistivity effects}},
  author = {Abrikosov, A. A.},
  journal = {Physics Physique Fizika},
  volume = {2},
  issue = {1},
  pages = {5--20},
  numpages = {16},
  year = {1965},
  month = {Sep},
  publisher = {American Physical Society},
  doi = {10.1103/PhysicsPhysiqueFizika.2.5},
  url = {https://link.aps.org/doi/10.1103/PhysicsPhysiqueFizika.2.5}
}

@ARTICLE{Lu16,
       author = {{Lu}, Yuan-Ming},
        title = "{Symmetric Z$_{2}$ spin liquids and their neighboring phases on triangular lattice}",
      journal = {\prb},
         year = 2016,
        month = apr,
       volume = {93},
       number = {16},
          eid = {165113},
        pages = {165113},
          doi = {10.1103/PhysRevB.93.165113},
archivePrefix = {arXiv},
       eprint = {1505.06495},
 primaryClass = {cond-mat.str-el},
       adsurl = {https://ui.adsabs.harvard.edu/abs/2016PhRvB..93p5113L}
}

@article{SSkagome,
  title = "{Kagome and triangular-lattice Heisenberg antiferromagnets: Ordering from quantum fluctuations and quantum-disordered ground states with unconfined bosonic spinons}",
  author = {Sachdev, Subir},
  journal = {Phys. Rev. B},
  volume = {45},
  issue = {21},
  pages = {12377--12396},
  numpages = {0},
  year = {1992},
  month = {Jun},
  publisher = {American Physical Society},
  doi = {10.1103/PhysRevB.45.12377},
  url = {https://link.aps.org/doi/10.1103/PhysRevB.45.12377}
}

@ARTICLE{Kaul08,
       author = {{Kaul}, Ribhu K. and {Sachdev}, Subir},
        title = "{Quantum criticality of U(1) gauge theories with fermionic and bosonic matter in two spatial dimensions}",
      journal = {\prb},
         year = 2008,
        month = apr,
       volume = {77},
       number = {15},
          eid = {155105},
        pages = {155105},
          doi = {10.1103/PhysRevB.77.155105},
archivePrefix = {arXiv},
       eprint = {0801.0723},
 primaryClass = {cond-mat.str-el},
       adsurl = {https://ui.adsabs.harvard.edu/abs/2008PhRvB..77o5105K}
}

@ARTICLE{Wietek24,
       author = {{Wietek}, Alexander and {Capponi}, Sylvain and {L{\"a}uchli}, Andreas M.},
        title = "{Quantum Electrodynamics in 2 +1 Dimensions as the Organizing Principle of a Triangular Lattice Antiferromagnet}",
      journal = {Physical Review X},
         year = 2024,
        month = apr,
       volume = {14},
       number = {2},
          eid = {021010},
        pages = {021010},
          doi = {10.1103/PhysRevX.14.021010},
archivePrefix = {arXiv},
       eprint = {2303.01585},
 primaryClass = {cond-mat.str-el},
       adsurl = {https://ui.adsabs.harvard.edu/abs/2024PhRvX..14b1010W}
}

@book{Polyakov_1987,
    author = "Polyakov, Alexander M.",
    title = "{Gauge Fields and Strings}",
    volume = "3",
    year = "1987",
    publisher="Taylor \& Francis",
    doi = {https://doi.org/10.1201/9780203755082}
}

@article{Christos_2024,
  title = {{Deconfined quantum criticality of nodal $d$-wave superconductivity, N\'eel order, and charge order on the square lattice at half-filling}},
  author = {Christos, Maine and Shackleton, Leyna and Sachdev, Subir and Luo, Zhu-Xi},
  journal = {Phys. Rev. Res.},
  volume = {6},
  issue = {3},
  pages = {033018},
  numpages = {25},
  year = {2024},
  month = {Jul},
  publisher = {American Physical Society},
  doi = {10.1103/PhysRevResearch.6.033018},
  url = {https://link.aps.org/doi/10.1103/PhysRevResearch.6.033018}
}

@article{Lu_2011,
  title = {{${\mathbb{Z}}_{2}$ spin liquids in the $S=\frac{1}{2}$ Heisenberg model on the kagome lattice: A projective symmetry-group study of Schwinger fermion mean-field states}},
  author = {Lu, Yuan-Ming and Ran, Ying and Lee, Patrick A.},
  journal = {Phys. Rev. B},
  volume = {83},
  issue = {22},
  pages = {224413},
  numpages = {11},
  year = {2011},
  month = {Jun},
  publisher = {American Physical Society},
  doi = {10.1103/PhysRevB.83.224413},
  url = {https://link.aps.org/doi/10.1103/PhysRevB.83.224413}
}

@article{Lu_2017,
  title = {{Unification of bosonic and fermionic theories of spin liquids on the kagome lattice}},
  author = {Lu, Yuan-Ming and Cho, Gil Young and Vishwanath, Ashvin},
  journal = {Phys. Rev. B},
  volume = {96},
  issue = {20},
  pages = {205150},
  numpages = {17},
  year = {2017},
  month = {Nov},
  publisher = {American Physical Society},
  doi = {10.1103/PhysRevB.96.205150},
  url = {https://link.aps.org/doi/10.1103/PhysRevB.96.205150}
}

@misc{Drescher-2025,
      title={{Spectral Functions of an Extended Antiferromagnetic $S=1/2$ Heisenberg Model on the Triangular Lattice}}, 
      author={Markus Drescher and Laurens Vanderstraeten and Roderich Moessner and Frank Pollmann},
      year={2025},
      eprint={2508.17292},
      archivePrefix={arXiv},
      primaryClass={cond-mat.str-el},
      url={https://arxiv.org/abs/2508.17292}, 
}

@article{Song-2019,
	author = {Song, Xue-Yang and Wang, Chong and Vishwanath, Ashvin and He, Yin-Chen},
	date = {2019/09/18},
	doi = {10.1038/s41467-019-11727-3},
	id = {Song2019},
	isbn = {2041-1723},
	journal = {Nat. Commun.},
	number = {1},
	pages = {4254},
	title = {{Unifying description of competing orders in two-dimensional quantum magnets}},
	url = {https://doi.org/10.1038/s41467-019-11727-3},
	volume = {10},
	year = {2019}}

@article{Song-2020,
  title = {{From Spinon Band Topology to the Symmetry Quantum Numbers of Monopoles in Dirac Spin Liquids}},
  author = {Song, Xue-Yang and He, Yin-Chen and Vishwanath, Ashvin and Wang, Chong},
  journal = {Phys. Rev. X},
  volume = {10},
  issue = {1},
  pages = {011033},
  numpages = {30},
  year = {2020},
  month = {Feb},
  publisher = {American Physical Society},
  doi = {10.1103/PhysRevX.10.011033},
  url = {https://link.aps.org/doi/10.1103/PhysRevX.10.011033}
}

@article{Iqbal-2016,
  title = {{Spin liquid nature in the Heisenberg ${J}_{1}\ensuremath{-}{J}_{2}$ triangular antiferromagnet}},
  author = {Iqbal, Yasir and Hu, Wen-Jun and Thomale, Ronny and Poilblanc, Didier and Becca, Federico},
  journal = {Phys. Rev. B},
  volume = {93},
  issue = {14},
  pages = {144411},
  numpages = {14},
  year = {2016},
  month = {Apr},
  publisher = {American Physical Society},
  doi = {10.1103/PhysRevB.93.144411},
  url = {https://link.aps.org/doi/10.1103/PhysRevB.93.144411}
}

@article{Zhu-2015,
  title = {{Spin liquid phase of the $S=\frac{1}{2}\phantom{\rule{4.pt}{0ex}}{J}_{1}\ensuremath{-}{J}_{2}$ Heisenberg model on the triangular lattice}},
  author = {Zhu, Zhenyue and White, Steven R.},
  journal = {Phys. Rev. B},
  volume = {92},
  issue = {4},
  pages = {041105},
  numpages = {4},
  year = {2015},
  month = {Jul},
  publisher = {American Physical Society},
  doi = {10.1103/PhysRevB.92.041105},
  url = {https://link.aps.org/doi/10.1103/PhysRevB.92.041105}
}

@article{Hu-2015,
  title = {{Competing spin-liquid states in the spin-$\frac{1}{2}$ Heisenberg model on the triangular lattice}},
  author = {Hu, Wen-Jun and Gong, Shou-Shu and Zhu, Wei and Sheng, D. N.},
  journal = {Phys. Rev. B},
  volume = {92},
  issue = {14},
  pages = {140403},
  numpages = {6},
  year = {2015},
  month = {Oct},
  publisher = {American Physical Society},
  doi = {10.1103/PhysRevB.92.140403},
  url = {https://link.aps.org/doi/10.1103/PhysRevB.92.140403}
}

@article{Hu-2019,
  title = {{Dirac Spin Liquid on the Spin-$1/2$ Triangular Heisenberg Antiferromagnet}},
  author = {Hu, Shijie and Zhu, W. and Eggert, Sebastian and He, Yin-Chen},
  journal = {Phys. Rev. Lett.},
  volume = {123},
  issue = {20},
  pages = {207203},
  numpages = {6},
  year = {2019},
  month = {Nov},
  publisher = {American Physical Society},
  doi = {10.1103/PhysRevLett.123.207203},
  url = {https://link.aps.org/doi/10.1103/PhysRevLett.123.207203}
}

@article{Budaraju-2025,
  title = {{Monopole excitations in the $U(1)$ Dirac spin liquid on the triangular lattice}},
  author = {Budaraju, Sasank and Parola, Alberto and Iqbal, Yasir and Becca, Federico and Poilblanc, Didier},
  journal = {Phys. Rev. B},
  volume = {111},
  issue = {12},
  pages = {125150},
  numpages = {10},
  year = {2025},
  month = {Mar},
  publisher = {American Physical Society},
  doi = {10.1103/PhysRevB.111.125150},
  url = {https://link.aps.org/doi/10.1103/PhysRevB.111.125150}
}

@article{Budaraju-2023,
  title = {{Piercing the Dirac spin liquid: From a single monopole to chiral states}},
  author = {Budaraju, Sasank and Iqbal, Yasir and Becca, Federico and Poilblanc, Didier},
  journal = {Phys. Rev. B},
  volume = {108},
  issue = {20},
  pages = {L201116},
  numpages = {6},
  year = {2023},
  month = {Nov},
  publisher = {American Physical Society},
  doi = {10.1103/PhysRevB.108.L201116},
  url = {https://link.aps.org/doi/10.1103/PhysRevB.108.L201116}
}

@misc{Feuerpfeil_2026,
      title={Unifying Dirac Spin Liquids on Square and Shastry-Sutherland Lattices via Fermionic Deconfined Criticality}, 
      author={Andreas Feuerpfeil and Leyna Shackleton and Atanu Maity and Ronny Thomale and Subir Sachdev and Yasir Iqbal},
      year={2026},
      eprint={2601.19980},
      archivePrefix={arXiv},
      primaryClass={cond-mat.str-el},
      url={https://arxiv.org/abs/2601.19980}, 
}

@article{Hermele2008,
  title = {Properties of an algebraic spin liquid on the kagome lattice},
  author = {Hermele, Michael and Ran, Ying and Lee, Patrick A. and Wen, Xiao-Gang},
  journal = {Phys. Rev. B},
  volume = {77},
  issue = {22},
  pages = {224413},
  numpages = {23},
  year = {2008},
  month = {Jun},
  publisher = {American Physical Society},
  doi = {10.1103/PhysRevB.77.224413},
  url = {https://link.aps.org/doi/10.1103/PhysRevB.77.224413}
}

@article{Rantner2002,
  title = {Spin correlations in the algebraic spin liquid: Implications for high-${T}_{c}$ superconductors},
  author = {Rantner, Walter and Wen, Xiao-Gang},
  journal = {Phys. Rev. B},
  volume = {66},
  issue = {14},
  pages = {144501},
  numpages = {20},
  year = {2002},
  month = {Oct},
  publisher = {American Physical Society},
  doi = {10.1103/PhysRevB.66.144501},
  url = {https://link.aps.org/doi/10.1103/PhysRevB.66.144501}
}

@misc{Maity-2026,
      title={{Unified gauge-theory description of quantum spin liquids on square-based frustrated lattices}}, 
      author={Atanu Maity and Andreas Feuerpfeil and Ronny Thomale and Subir Sachdev and Yasir Iqbal},
      year={2026},
      eprint={2603.15745},
      archivePrefix={arXiv},
      primaryClass={cond-mat.str-el},
      url={https://arxiv.org/abs/2603.15745}, 
}

@article{Xia2025,
	author = {Xia, Yiyu and Han, Zhongdong and Watanabe, Kenji and Taniguchi, Takashi and Shan, Jie and Mak, Kin Fai},
	date = {2025/01/01},
	doi = {10.1038/s41586-024-08116-2},
	id = {Xia2025},
	isbn = {1476-4687},
	journal = {Nature (London)},
	number = {8047},
	pages = {833--838},
	title = {{Superconductivity in twisted bilayer WSe$_2$}},
	url = {https://doi.org/10.1038/s41586-024-08116-2},
	volume = {637},
	year = {2025}}

@article{Xia2026,
	author = {Xia, Yiyu and Han, Zhongdong and Zhu, Jiacheng and Zhang, Yichi and Kn{\"u}ppel, Patrick and Watanabe, Kenji and Taniguchi, Takashi and Mak, Kin Fai and Shan, Jie},
	date = {2026/02/01},
	doi = {10.1038/s41586-025-10049-3},
	id = {Xia2026},
	isbn = {1476-4687},
	journal = {Nature},
	number = {8102},
	pages = {585--591},
	title = "{Bandwidth-tuned Mott transition and superconductivity in moir{\'e} WSe$_2$}",
	url = {https://doi.org/10.1038/s41586-025-10049-3},
	volume = {650},
	year = {2026}}

@article{Iqbal-2011_vbc,
  title = {{Valence-bond crystal in the extended kagome spin-$\frac{1}{2}$ quantum Heisenberg antiferromagnet: A variational Monte Carlo approach}},
  author = {Iqbal, Yasir and Becca, Federico and Poilblanc, Didier},
  journal = {Phys. Rev. B},
  volume = {83},
  issue = {10},
  pages = {100404(R)},
  numpages = {4},
  year = {2011},
  month = {Mar},
  publisher = {American Physical Society},
  doi = {10.1103/PhysRevB.83.100404},
  url = {https://link.aps.org/doi/10.1103/PhysRevB.83.100404}
}

@misc{zhang2026,
      title={{Large-flavor route to a stable U(1) Dirac spin liquid on the maple-leaf lattice}}, 
      author={Yunchao Zhang and Andreas Feuerpfeil and Subir Sachdev and Ronny Thomale and Yasir Iqbal},
      year={2026},
      eprint={2605.21587},
      archivePrefix={arXiv},
      primaryClass={cond-mat.str-el},
      url={https://arxiv.org/abs/2605.21587}, 
}

@article{
Christos_2023,
author = {Maine Christos  and Zhu-Xi Luo  and Henry Shackleton  and Ya-Hui Zhang  and Mathias S. Scheurer  and Subir Sachdev },
title = {A model of d-wave superconductivity, antiferromagnetism, and charge order on the square lattice},
journal = {Proceedings of the National Academy of Sciences},
volume = {120},
number = {21},
pages = {e2302701120},
year = {2023},
doi = {10.1073/pnas.2302701120},
URL = {https://www.pnas.org/doi/abs/10.1073/pnas.2302701120},
eprint = {https://www.pnas.org/doi/pdf/10.1073/pnas.2302701120},
}

@article{Iqbal_2012,
doi = {10.1088/1367-2630/14/11/115031},
url = {https://doi.org/10.1088/1367-2630/14/11/115031},
year = {2012},
month = {nov},
publisher = {IOP Publishing},
volume = {14},
number = {11},
pages = {115031},
author = {Iqbal, Yasir and Becca, Federico and Poilblanc, Didier},
title = {{Valence-bond crystals in the kagome spin-$1/2$ Heisenberg antiferromagnet: a symmetry classification and projected wave function study}},
journal = {New J. Phys.}
}

@article{Kim2024,
	author = {Kim, Sunghoon and Mendez-Valderrama, Juan Felipe and Wang, Xuepeng and Chowdhury, Debanjan},
	date = {2025/02/17},
	doi = {10.1038/s41467-025-56816-8},
	id = {Kim2025},
	isbn = {2041-1723},
	journal = {Nat. Commun.},
	number = {1},
	pages = {1701},
	title = {{Theory of correlated insulators and superconductor at $\nu=1$ in twisted WSe$_2$}},
	url = {https://doi.org/10.1038/s41467-025-56816-8},
	volume = {16},
	year = {2025}}

@article{Peng2025,
  title = {{Itinerant magnetism in twisted bilayer ${\mathrm{WSe}}_{2}$ and ${\mathrm{MoTe}}_{2}$}},
  author = {Peng, Liangtao and De Beule, Christophe and Lai, Yiyang and Li, Du and Yang, Li and Mele, E. J. and Adam, Shaffique},
  journal = {Phys. Rev. B},
  volume = {113},
  issue = {16},
  pages = {L161119},
  numpages = {6},
  year = {2026},
  month = {Apr},
  publisher = {American Physical Society},
  doi = {10.1103/hmgc-shx3},
  url = {https://link.aps.org/doi/10.1103/hmgc-shx3}
}

@article{Iqbal-2011,
  title = {{Projected wave function study of ${\mathbb{Z}}_{2}$ spin liquids on the kagome lattice for the spin-$\frac{1}{2}$ quantum Heisenberg antiferromagnet}},
  author = {Iqbal, Yasir and Becca, Federico and Poilblanc, Didier},
  journal = {Phys. Rev. B},
  volume = {84},
  issue = {2},
  pages = {020407},
  numpages = {4},
  year = {2011},
  month = {Jul},
  publisher = {American Physical Society},
  doi = {10.1103/PhysRevB.84.020407},
  url = {https://link.aps.org/doi/10.1103/PhysRevB.84.020407}
}

@article{Jiang-2026,
  title = {Competing states in the S = 1/2 triangular-lattice ${J}_{1}$-${J}_{2}$ Heisenberg model: A dynamical density-matrix renormalization group study},
  author = {Jiang, Shengtao and White, Steven R. and Kivelson, Steven A. and Jiang, Hong-Chen},
  journal = {Phys. Rev. Lett.},
  pages = {},
  year = {2026},
  month = {Jun},
  publisher = {American Physical Society},
  doi = {10.1103/zmnz-tkq2},
  url = {https://link.aps.org/doi/10.1103/zmnz-tkq2}
}

@misc{Kovalska-2026,
      title={{Revisiting the $J_1$-$J_2$ Heisenberg Model on a Triangular Lattice: Quasi-Degenerate Ground States and Phase Competition}}, 
      author={Oleksandra Kovalska and Ester Pagès Fontanella and Benedikt Schneider and Hong-Hao Tu and Jan von Delft},
      year={2026},
      eprint={2603.08650},
      archivePrefix={arXiv},
      primaryClass={cond-mat.str-el},
      url={https://arxiv.org/abs/2603.08650}, 
}

@article{Mei-2017,
  title = {{Gapped spin liquid with ${\mathbb{Z}}_{2}$ topological order for the kagome Heisenberg model}},
  author = {Mei, Jia-Wei and Chen, Ji-Yao and He, Huan and Wen, Xiao-Gang},
  journal = {Phys. Rev. B},
  volume = {95},
  issue = {23},
  pages = {235107},
  numpages = {9},
  year = {2017},
  month = {Jun},
  publisher = {American Physical Society},
  doi = {10.1103/PhysRevB.95.235107},
  url = {https://link.aps.org/doi/10.1103/PhysRevB.95.235107}
}

@article{Iqbal-2013,
  title = {{Gapless spin-liquid phase in the kagome spin-$\frac{1}{2}$ Heisenberg antiferromagnet}},
  author = {Iqbal, Yasir and Becca, Federico and Sorella, Sandro and Poilblanc, Didier},
  journal = {Phys. Rev. B},
  volume = {87},
  issue = {6},
  pages = {060405},
  numpages = {5},
  year = {2013},
  month = {Feb},
  publisher = {American Physical Society},
  doi = {10.1103/PhysRevB.87.060405},
  url = {https://link.aps.org/doi/10.1103/PhysRevB.87.060405}
}

@article{Iqbal-2015,
  title = {{Spin-$\frac{1}{2}$ Heisenberg ${J}_{1}\text{\ensuremath{-}}{J}_{2}$ antiferromagnet on the kagome lattice}},
  author = {Iqbal, Yasir and Poilblanc, Didier and Becca, Federico},
  journal = {Phys. Rev. B},
  volume = {91},
  issue = {2},
  pages = {020402},
  numpages = {5},
  year = {2015},
  month = {Jan},
  publisher = {American Physical Society},
  doi = {10.1103/PhysRevB.91.020402},
  url = {https://link.aps.org/doi/10.1103/PhysRevB.91.020402}
}

@article{Iqbal-2016_breathing,
  title = {{Persistence of the gapless spin liquid in the breathing kagome Heisenberg antiferromagnet}},
  author = {Iqbal, Yasir and Poilblanc, Didier and Thomale, Ronny and Becca, Federico},
  journal = {Phys. Rev. B},
  volume = {97},
  issue = {11},
  pages = {115127},
  numpages = {8},
  year = {2018},
  month = {Mar},
  publisher = {American Physical Society},
  doi = {10.1103/PhysRevB.97.115127},
  url = {https://link.aps.org/doi/10.1103/PhysRevB.97.115127}
}

@article{Ferrari-2019,
  title = {{Dynamical Structure Factor of the ${J}_{1}\ensuremath{-}{J}_{2}$ Heisenberg Model on the Triangular Lattice: Magnons, Spinons, and Gauge Fields}},
  author = {Ferrari, Francesco and Becca, Federico},
  journal = {Phys. Rev. X},
  volume = {9},
  issue = {3},
  pages = {031026},
  numpages = {12},
  year = {2019},
  month = {Aug},
  publisher = {American Physical Society},
  doi = {10.1103/PhysRevX.9.031026},
  url = {https://link.aps.org/doi/10.1103/PhysRevX.9.031026}
}

@ARTICLE{Kanoda18,
       author = {{Furukawa}, Tetsuya and {Kobashi}, Kazuhiko and {Kurosaki}, Yosuke and {Miyagawa}, Kazuya and {Kanoda}, Kazushi},
        title = "{Quasi-continuous transition from a Fermi liquid to a spin liquid in {\ensuremath{\kappa}}-(ET)$_{2}$Cu$_{2}$(CN)$_{3}$}",
      journal = {Nature Communications},
         year = 2018,
        month = jan,
       volume = {9},
          eid = {307},
        pages = {307},
          doi = {10.1038/s41467-017-02679-7},
archivePrefix = {arXiv},
       eprint = {1707.05586},
 primaryClass = {cond-mat.str-el},
       adsurl = {https://ui.adsabs.harvard.edu/abs/2018NatCo...9..307F}
}

@article{Kanoda24,
author = {Oike, Hiroshi and Taniguchi, Hiromi and Miyagawa, Kazuya and Kanoda, Kazushi},
title = "{Mottness and Spin Liquidity in a Doped Organic Superconductor $\kappa$-(BEDT-TTF)$_4$Hg$_{2.89}$Br$_8$}",
journal = {Journal of the Physical Society of Japan},
volume = {93},
number = {4},
pages = {042001},
year = {2024},
doi = {10.7566/JPSJ.93.042001}
}

@ARTICLE{Sengupta25,
       author = {Duri{\'c}, Tanja and {Chung}, Jia Hui and {Yang}, Bo and {Sengupta}, Pinaki},
        title = "{Spin-1/2 Kagome Heisenberg Antiferromagnet: Machine Learning Discovery of the Spinon Pair-Density-Wave Ground State}",
      journal = {Physical Review X},
         year = 2025,
        month = jan,
       volume = {15},
       number = {1},
          eid = {011047},
        pages = {011047},
          doi = {10.1103/PhysRevX.15.011047},
archivePrefix = {arXiv},
       eprint = {2401.02866},
 primaryClass = {cond-mat.str-el},
       adsurl = {https://ui.adsabs.harvard.edu/abs/2025PhRvX..15a1047D}
}

@article{Sherman-2023,
  title = {{Spectral function of the ${J}_{1}\ensuremath{-}{J}_{2}$ Heisenberg model on the triangular lattice}},
  author = {Sherman, Nicholas E. and Dupont, Maxime and Moore, Joel E.},
  journal = {Phys. Rev. B},
  volume = {107},
  issue = {16},
  pages = {165146},
  numpages = {19},
  year = {2023},
  month = {Apr},
  publisher = {American Physical Society},
  doi = {10.1103/PhysRevB.107.165146},
  url = {https://link.aps.org/doi/10.1103/PhysRevB.107.165146}
}

@article{Iqbal-2021,
  title = {{Gutzwiller projected states for the ${J}_{1}\ensuremath{-}{J}_{2}$ Heisenberg model on the Kagome lattice: Achievements and pitfalls}},
  author = {Iqbal, Yasir and Ferrari, Francesco and Chauhan, Aishwarya and Parola, Alberto and Poilblanc, Didier and Becca, Federico},
  journal = {Phys. Rev. B},
  volume = {104},
  issue = {14},
  pages = {144406},
  numpages = {9},
  year = {2021},
  month = {Oct},
  publisher = {American Physical Society},
  doi = {10.1103/PhysRevB.104.144406},
  url = {https://link.aps.org/doi/10.1103/PhysRevB.104.144406}
}

@ARTICLE{Tennant24,
       author = {{Scheie}, A.~O. and {Lee}, Minseong and {Wang}, Kevin and {Laurell}, P. and {Choi}, E.~S. and {Pajerowski}, D. and {Zhang}, Qingming and {Ma}, Jie and {Zhou}, H.~D. and {Lee}, Sangyun and {Huan}, Chao and {Thomas}, S.~M. and {Ajeesh}, M.~O. and {Rosa}, P.~F.~S. and {Chen}, Ao and {Zapf}, Vivien S. and {Heyl}, M. and {Batista}, C.~D. and {Dagotto}, E. and {Moore}, J.~E. and {Tennant}, D. Alan},
        title = "{Spectrum and low-temperature bulk properties of triangular quantum spin liquid candidate NaYbSe$_2$}",
      journal = {arXiv e-prints},
         year = 2024,
        month = jun,
          eid = {arXiv:2406.17773},
        pages = {arXiv:2406.17773},
          doi = {10.48550/arXiv.2406.17773},
archivePrefix = {arXiv},
       eprint = {2406.17773},
 primaryClass = {cond-mat.str-el},
       adsurl = {https://ui.adsabs.harvard.edu/abs/2024arXiv240617773S}
}

@article{chinese_na,
author = {Jia, Ya-Ting  and Gong, Chun-Sheng and Liu, Yi-Xuan and Zhao, Jian-Fa and Dong, Cheng and Dai, Guang-Yang and Li, Xiao-Dong and Lei, He-Chang and Yu, Run-Ze and Zhang, Guang-Ming and Jin, Chang-Qing},
title = "{Mott Transition and Superconductivity in Quantum Spin Liquid Candidate NaYbSe$_2$}",
journal = {Chinese Physics Letters},
volume = {37},
pages = {097404},
year = {2020},
doi = {10.1088/0256-307X/37/9/097404}
}

@ARTICLE{TSSSMV03,
   author = {{Senthil}, T. and {Sachdev}, S. and {Vojta}, M.},
    title = "{Fractionalized Fermi Liquids}",
  journal = {Phys. Rev. Lett.},
   eprint = {cond-mat/0209144},
     year = 2003,
    month = may,
   volume = 90,
   number = 21,
      eid = {216403},
    pages = {216403},
      doi = {10.1103/PhysRevLett.90.216403},
   adsurl = {http://adsabs.harvard.edu/abs/2003PhRvL..90u6403S}
}

@article{Iqbal-2014_gap,
  title = {{Vanishing spin gap in a competing spin-liquid phase in the kagome Heisenberg antiferromagnet}},
  author = {Iqbal, Yasir and Poilblanc, Didier and Becca, Federico},
  journal = {Phys. Rev. B},
  volume = {89},
  issue = {2},
  pages = {020407(R)},
  numpages = {5},
  year = {2014},
  month = {Jan},
  publisher = {American Physical Society},
  doi = {10.1103/PhysRevB.89.020407},
  url = {https://link.aps.org/doi/10.1103/PhysRevB.89.020407}
}

@article{TSMVSS04,
   author = {{Senthil}, T. and {Vojta}, M. and {Sachdev}, S.},
    title = "{Weak magnetism and non-Fermi liquids near heavy-fermion critical points}",
  journal = {Phys. Rev. B},
   eprint = {cond-mat/0305193},
     year = 2004,
    month = jan,
   volume = 69,
   number = 3,
      eid = {035111},
    pages = {035111},
      doi = {10.1103/PhysRevB.69.035111},
   adsurl = {http://adsabs.harvard.edu/abs/2004PhRvB..69c5111S}
}

@ARTICLE{IvanovSenthil,
       author = {{Ivanov}, D.~A. and {Senthil}, T.},
        title = "{Projected wave functions for fractionalized phases of quantum spin systems}",
      journal = {Phys. Rev. B},
         year = 2002,
        month = sep,
       volume = {66},
       number = {11},
          eid = {115111},
        pages = {115111},
          doi = {10.1103/PhysRevB.66.115111},
archivePrefix = {arXiv},
       eprint = {cond-mat/0204043},
 primaryClass = {cond-mat.str-el},
       adsurl = {https://ui.adsabs.harvard.edu/abs/2002PhRvB..66k5111I}
}

@article{Boyack_2018,
  title = {Transition between algebraic and $Z_2$ quantum spin liquids at large $N$},
  author = {Boyack, Rufus and Lin, Chien-Hung and Zerf, Nikolai and Rayyan, Ahmed and Maciejko, Joseph},
  journal = {Phys. Rev. B},
  volume = {98},
  issue = {3},
  pages = {035137},
  numpages = {11},
  year = {2018},
  month = {Jul},
  publisher = {American Physical Society},
  doi = {10.1103/PhysRevB.98.035137},
  url = {https://link.aps.org/doi/10.1103/PhysRevB.98.035137}
}

@misc{Dumitrescu_2026,
      title="{From QED$_3$ to Self-Dual Multicriticality in the Fradkin-Shenker Model}", 
      author={Thomas T. Dumitrescu and Pierluigi Niro and Ryan Thorngren},
      year={2026},
      eprint={2602.23420},
      archivePrefix={arXiv},
      primaryClass={cond-mat.str-el},
      url={https://arxiv.org/abs/2602.23420}, 
}

@misc{Ebert2026a,
      title={{Competing Paramagnetic Phases in the Maple-Leaf Heisenberg Antiferromagnet}}, 
      author={Paul L. Ebert and Yasir Iqbal and Alexander Wietek},
      year={2026},
      eprint={2601.05308},
      archivePrefix={arXiv},
      primaryClass={cond-mat.str-el},
      url={https://arxiv.org/abs/2601.05308}, 
}

@misc{Ebert2026b,
      title={{Incommensurate Spin-Density Waves in a Frustrated Maple-Leaf Lattice Ferromagnet}}, 
      author={Paul L. Ebert and Yasir Iqbal and Alexander Wietek},
      year={2026},
      eprint={2605.12592},
      archivePrefix={arXiv},
      primaryClass={cond-mat.str-el},
      url={https://arxiv.org/abs/2605.12592}, 
}

@article{Gresista2026,
url = {https://doi.org/10.1515/zna-2025-0376},
title = {Unconventional orders in the maple-leaf ferro-antiferromagnetic Heisenberg model},
title = {},
author = {Lasse Gresista and Dominik Kiese and Simon Trebst and Yasir Iqbal},
pages = {433--452},
volume = {81},
number = {6},
journal = {Zeitschrift für Naturforschung A},
doi = {doi:10.1515/zna-2025-0376},
year = {2026},
lastchecked = {2026-07-23}
}

@misc{Schmoll2026,
      title={{An extended non-magnetic phase in the spin-1/2 Heisenberg antiferromagnet from the ruby to the maple-leaf lattice}}, 
      author={Philipp Schmoll and Jan Naumann and Erik Lennart Weerda and Jens Eisert and Yasir Iqbal},
      year={2026},
      eprint={2407.07145},
      archivePrefix={arXiv},
      primaryClass={cond-mat.str-el},
      url={https://arxiv.org/abs/2407.07145}, 
}

@article{Sonnenschein-2024,
  title = {{Candidate quantum spin liquids on the maple-leaf lattice}},
  author = {Sonnenschein, Jonas and Maity, Atanu and Liu, Chunxiao and Thomale, Ronny and Ferrari, Francesco and Iqbal, Yasir},
  journal = {Phys. Rev. B},
  volume = {110},
  issue = {1},
  pages = {014414},
  numpages = {27},
  year = {2024},
  month = {Jul},
  publisher = {American Physical Society},
  doi = {10.1103/PhysRevB.110.014414},
  url = {https://link.aps.org/doi/10.1103/PhysRevB.110.014414}
}

@article{Gresista2023,
  title = {{Candidate quantum disordered intermediate phase in the Heisenberg antiferromagnet on the maple-leaf lattice}},
  author = {Gresista, Lasse and Hickey, Ciar\'an and Trebst, Simon and Iqbal, Yasir},
  journal = {Phys. Rev. B},
  volume = {108},
  issue = {24},
  pages = {L241116},
  numpages = {7},
  year = {2023},
  month = {Dec},
  publisher = {American Physical Society},
  doi = {10.1103/PhysRevB.108.L241116},
  url = {https://link.aps.org/doi/10.1103/PhysRevB.108.L241116}
}

@article{Beck-2024,
  title = {{Phase diagram of the $J\text{\ensuremath{-}}{J}_{d}$ Heisenberg model on the maple leaf lattice: Neural networks and density matrix renormalization group}},
  author = {Beck, Jonas and Bodky, Jonathan and Motruk, Johannes and M\"uller, Tobias and Thomale, Ronny and Ghosh, Pratyay},
  journal = {Phys. Rev. B},
  volume = {109},
  issue = {18},
  pages = {184422},
  numpages = {10},
  year = {2024},
  month = {May},
  publisher = {American Physical Society},
  doi = {10.1103/PhysRevB.109.184422},
  url = {https://link.aps.org/doi/10.1103/PhysRevB.109.184422}
}

\end{document}